\documentclass[aj,twocolumn]{aastex61}
\pdfoutput=1
\usepackage{amsmath,amstext}
\usepackage[T1]{fontenc}
\usepackage[utf8]{inputenc}
\usepackage{apjfonts} 
\usepackage[figure,figure*]{hypcap}
\usepackage{amsmath,amssymb,amsfonts,graphicx}
\usepackage{todonotes}
\usepackage{color}
\usepackage{hyperref}
\usepackage{graphicx}

\usepackage{todonotes}   

\newcommand{\Msun}{{\rm M_{\odot}}}

\newcommand{\degree}{\ensuremath{^\circ}}

\shorttitle{A NIR view of NGC\,6300}
\shortauthors{Gaspar et al.}

\begin{document}

\title{A Near Infrared View of Nearby Galaxies: The Case of NGC\,6300 }

\author{Gaspar G.}
\affiliation{Observatorio Astron\'omico de C\'ordoba, Laprida 854, X5000BGR, C\'ordoba, Argentina.}
\affiliation{Consejo de Investigaciones Cient\'{i}ficas y T\'ecnicas de la Rep\'ublica Argentina, Avda. Rivadavia 1917, C1033AAJ, CABA, Argentina.}

\author{D\'{i}az R.J.}
\affiliation{Observatorio Astron\'omico de C\'ordoba, Laprida 854, X5000BGR, C\'ordoba, Argentina.}
\affiliation{Consejo de Investigaciones Cient\'{i}ficas y T\'ecnicas de la Rep\'ublica Argentina, Avda. Rivadavia 1917, C1033AAJ, CABA, Argentina.}
\affiliation{Gemini Observatory, 950 N Cherry Ave., Tucson, AZ85719, USA.}

\author{Mast D.}
\affiliation{Observatorio Astron\'omico de C\'ordoba, Laprida 854, X5000BGR, C\'ordoba, Argentina.}
\affiliation{Consejo de Investigaciones Cient\'{i}ficas y T\'ecnicas de la Rep\'ublica Argentina, Avda. Rivadavia 1917, C1033AAJ, CABA, Argentina.}

\author{D'Ambra A.}
\affiliation{Observatorio Astron\'omico de C\'ordoba, Laprida 854, X5000BGR, C\'ordoba, Argentina.}

\author{Ag\"{u}ero M.P.}
\affiliation{Observatorio Astron\'omico de C\'ordoba, Laprida 854, X5000BGR, C\'ordoba, Argentina.}
\affiliation{Consejo de Investigaciones Cient\'{i}ficas y T\'ecnicas de la Rep\'ublica Argentina, Avda. Rivadavia 1917, C1033AAJ, CABA, Argentina.}

\author{G\"{u}nthardt G.}
\affiliation{Observatorio Astron\'omico de C\'ordoba, Laprida 854, X5000BGR, C\'ordoba, Argentina.}

\begin{abstract}

We present a near-infrared study of the Seyfert 2 galaxy NGC\,6300, based on subarcsecond images and long slit spectroscopy obtained with Flamingos-2 at Gemini South. We have found that the peak of the nuclear continuum emission in the $K_s$ band and the surrounding nuclear disk are 25\,pc off-center with respect to the center of symmetry of the larger scale circumnuclear disk, suggesting that this black hole is still not fixed in the galaxy potential well. The molecular gas radial velocity curve yields a central black hole upper mass estimation of $M_{SMBH}^{upper}=(6\pm 2) \times 10^{7}\,\Msun$. The Pa$\beta$ emission line has a strongly asymmetric profile with a blueshifted broad component that we associate with a nuclear ionized gas outflow. We have found in the $K_s$-band spectra that the slope of the continuum becomes steeper with increasing radii, which can be explained as the presence of large amounts of hot dust not only in the nucleus but also in the circumnuclear region up to $r=27$\,pc. In fact, the nuclear red excess obtained after subtracting the stellar contribution resembles to that of a blackbody with temperatures around 1200\,K. This evidence supports the idea that absorbing material located around the nucleus, but not close enough to be the torus of the unified model, could be responsible for at least part of the nuclear obscuration in this Seyfert 2 nucleus.

\end{abstract}


\section{Introduction}
Active Galactic Nuclei are among the most intriguing objects because they are present throughout most of the universe's history, meaning they must be an essential ingredient in the galactic and cosmic evolution, yet there are several aspects that are still not completely understood, such as, for example, AGN feedback \citep{Fabian2012} or the M$_{SMBH}-\sigma$ relation \citep{Tremaine2002}. It is now accepted that almost all massive enough galaxies harbor a super-massive black hole (SMBH) in their nuclei. This has the strong implication that all massive galaxies should have gone through an AGN phase in some point of their lives. This phase could be episodic or long-lasting depending on galaxy environment and fuel supply \citep{Shankar2009}. These duty cycles and feeding mechanisms are not related to a particular kind of galaxy, but are intrinsic to galaxy evolution as a whole.

The unified model (UM) is to-date the most accepted model to understand different types of active nuclei. It proposes that it is actually a single phenomenon that is being observed from different directions and that the observed properties are due to the anisotropy of the object \citep{Blandford1978,Antonucci1993}. According to the model, the object consists of a SMBH surrounded by an accretion disk, gas clouds and a torus-shaped structure composed of colder molecular gas and dust. In this scheme, the torus is responsible for the high anisotropy of the AGN emission because it acts like a shield to the emission of the central source when seen from low inclination directions.

In the last few decades, some works have emerged presenting evidence that the panorama is more complex than a simple orientation issue, proposing different geometries for the dusty absorber \citep{Nenkova2002,Dullemond2005,Elitzur2006,AlonsoHerrero2011,RamosAlmeida2011,Stalevski2012,Elitzur2012}  and giving a central role to the host galaxy interstellar medium (ISM)  to explain the differences that arise while observing active nuclei \citep{Malkan1998,Matt2000,Reunanen2003,Goulding2009,Goulding2012}. One thing is certain, the very nuclear dust structure is not well constrained yet nor the feeding mechanism that has driven AGN growth through cosmic times until the present.

Also, the mass of the central SMBH could play an important role in the emission and geometry we observe. A low-mass black hole will still not be fixed in the gravitational potential well of the galaxy \citep{Emsellem2015}, in which case it would wobble around the galaxy gravitational center. This can be observed as a break in the symmetry between the global structure of the galaxy and the central emission source (e.g. \citealt{Arribas1994}, \citealt{Arribas1996}, \citealt{Diaz99,Diaz2006}, \citealt{Mast06}, and \citealt{Gunthardt2015}), and could be part of the mechanism that drives material from the last few hundred parsecs onto accretion in the SMBH.

Much of the uncertainty around the UM could arise from the fact that historically AGNs have been mostly studied and classified in the optical band. In the last two decades studies in other bands have become frequent \citep{Polletta1996,Xu1999,Greusard2000,Beckmann2006,Peng2006,Munoz2007,Lacy2013,She2017}. The optical band is the most affected by dust extinction in our galaxy, in the host galaxy ISM, and in the very nucleus, and therefore it is expected that observations in other bands show higher emission levels than in the optical for the most obscured nuclei. In fact, optical surveys miss half of the AGNs in comparison with the MIR \citep{Goulding2009}, and misclassification of the emission type occurs when studying AGNs in the optical alone \citep{Nagar2002,Riffel2006,RamosAlmeida2008}.

The near-infrared (NIR) bands offer the advantage of lowering the galactic extinction both in the Milky Way and in the host galaxy, reaching 10\% of the $V$-band extinction in the $K_s$ band. This allows NIR photons to escape easily and provides a more detailed view of the structure and physical state of the circumnuclear regions. At the same time, more energetic photons absorbed by dust are re-emitted in the infrared allowing us to map the most dusty regions. In particular, in the $K$ band is possible to observe a red excess associated with hot dust in the nuclear regions (e.g. \citealt{Thatte1997} \citealt{Alonso-Herrero1998}, \citealt{Ferruit2004}, \citealt{MullerSanchez2006},\citealt{RamosAlmeida2009}, \citealt{MullerSanchez2018}). NIR also has the advantage of being observable from the ground and now we count with several facilities optimized for imaging and spectroscopy in the NIR bands. For this work we used Flamingos-2 (F2), an imager and spectrometer mounted on the 8.1\,m Gemini South telescope that achieves observations with subarcsecond resolutions. This implies spatial scales under 100\,pc for galaxies in the nearby universe ($\lesssim 21$\, Mpc or $z\lesssim 0.0052$ for $H_0 =75$ km  s$^{-1}$  Mpc$^{-1}$).

We are observing a sample of nearby galaxies in image and spectroscopic modes with F2 to constrain the geometrical and dust properties of the nuclear regions of galaxies with a variety of emission types including several obscured AGNs, and link them with large-scale properties of the host galaxies. 
Here we present results for NGC\,6300 (d = 12.26 Mpc, z=0.0037), a ringed barred galaxy of the southern hemisphere that host an active nucleus identified as a Seyfert 2 in the optical by \cite{Phillips1983}, and identified as well in the X-rays by \cite{Awaki2005}, in the mid-infrared by \cite{Goulding2009}, and in radio frequencies by \citep{Morganti1999}. It presents prominent dust lanes that reach the nuclear region, probably causing strong obscuration in the host galaxy at central and larger scales (see Fig.\ref{Kimage}), making it a good candidate to explore the ISM role in the AGN emission using infrared data. 

Pioneering works on NGC\,6300 was made in the 80-90's. \cite{Phillips1983} studied the optical spectrum and classified it as a Seyfert 2 using the BPT diagram. \cite{Buta1987} presented a photometric and kinematic study using BV surface photometry and emission line radial velocities. \cite{Ryder1996} presented a kinematic study based in HI observations where they linked the galaxy ring with a Lindblad resonance and derived the bar pattern speed.

In the last few decades, a series of X-ray observations focused on its nucleus. The central source was found in a Compton-thick state by \textit{RXTE} in 1999 \citep{Leighly1999} and then in a Compton-thin state by \textit{BeppoSAX} \citep{Guainazzi2002} and \textit{XMM-Newton} \citep{Matsumoto2004}. There is still no consensus whether these variations are common in Seyfert 2 nuclei or not \citep{Risaliti2002,Elvis2004,Puccetti2004,Bianchi2005}. Furthermore, their origin is not well understood yet in the framework of the UM, but it may be related with the very geometry of the circumnuclear region and, in particular, of the absorbing material \citep{Matt2000,Risaliti2005}.

In this paper we present the analysis of Flamingos-2 imaging and spectra in the $J$ and $K_{s}$ bands for NGC\,6300. In Section \ref{obs} the observation and data reduction process are described. In Section \ref{results} the results are presented starting by the imaging data in Section \ref{structure} and following with the spectral data analysis in Section \ref{spectra}. The results are discussed in Section \ref{results} and the summary and conclusions are presented in Section \ref{summary}.

\section{Observations and Data Reduction} \label{obs}

Longslit spectra and images of NGC\,6300 were taken with Flamingos-2 \citep{Eikenberry2012,Diaz2013} at the 8.1\,m Gemini South Telescope during a commissioning run in 2013 April 29 and June 24. The broadband filters $J$, $H$ and $K_s$ available for F2 were used in combination with the R3K grism and the 3-pixel wide (0\farcs54) long slit, which resulted in a nominal resolving power of $\sim2000$ at the center of the $K_s$ band spectral range ($2.157 \mu$m) dropping to $\sim1000$ at the ends, as measured from the OH sky emission lines. In this paper we present $J$- and $K_s$-band results from a total of 4.5 hs. observations. Considering the sky subtraction overheads and telluric standards, this yielded on-source total exposure times of 1800 sec in $J$ and 3600 sec in $K_s$. The observations were performed following an ABBA sequence. The slit was positioned along the semi-major axis at $PA=128.5\degree$ avoiding the light from two foreground stars near the nucleus of NGC\,6300 (see Fig.\ref{Kimage}). The use of the on-instrument wavefront sensor (OIWFS) allowed a precision of half pixel (0\farcs09) in slit positioning and guiding. The OIWFS guide star counts were consistent with photometric observing conditions.

We used a standard reduction procedure using the Gemini IRAF\footnote{http://iraf.noao.edu/} data reduction package\footnote{http://www.gemini.edu/sciops/data-and-results/processing-software}, performing dark subtraction, flat-fielding, sky emission subtraction, wavelength calibration and telluric correction. All steps except the wavelength calibration and the telluric correction were performed following the Flamingos-2 tasks. The wavelength calibration was performed step by step using the standard PyRAF tasks {\it identify}, {\it refspectra}, and {\it dispcord} applied in the usual way. This procedure lowered the calibration uncertainties with respect to the obtained using the {\it nswavelength} task. The telluric correction was carried out also with PyRAF\footnote{http://www.stsci.edu/institute/software\_hardware/pyraf/}, first generating a synthetic continuum corresponding to the spectral type of the telluric star, dividing the star spectrum by the template and then manually subtracting the star intrinsic absorptions.

We performed extractions from the two-dimensional spectrum to study the radial distribution of the spectroscopic properties in the circumnuclear region of NGC6300, with the following scheme: the nuclear core extraction is 0\farcs18 wide in order to minimize the contribution from off-nuclear regions; then we extracted three spectra at each side of the nucleus with an extraction aperture of 0\farcs54 taking into account the spatial resolution.

We have also obtained images in $J$, $H$ and $K_s$ filters from which we derived an average image quality of 0\farcs45. This, in combination with the F2 pixel scale of 0\farcs18 pixel$^{-1}$ gives a spacial resolution of 33 pc at the galaxy mean distance of 12.26\,Mpc (from NED\footnote{https://ned.ipac.caltech.edu/}). 
The images were partially reduced using, additionally, the data reduction package  designed for the instrument. Then, the coadding was done manually giving more weight to the central 2 arcmin of the image, in order to maximize the spatial resolution in the central region of the galaxy and to avoid the effect of optical distortions in the outer radii of the images in the coadded result.

We have flux calibrated the images using the 2MASS stars' magnitudes in the field, $J$, $H$, $K_s$, subtracting the contribution of the galaxy's background. In this way, zero points were calculated.

\section{Results and Discussion}\label{results}

\subsection{The nuclear region structure}\label{structure}

\begin{figure*}[!t] 
  \centering
  \includegraphics[width=1\textwidth]{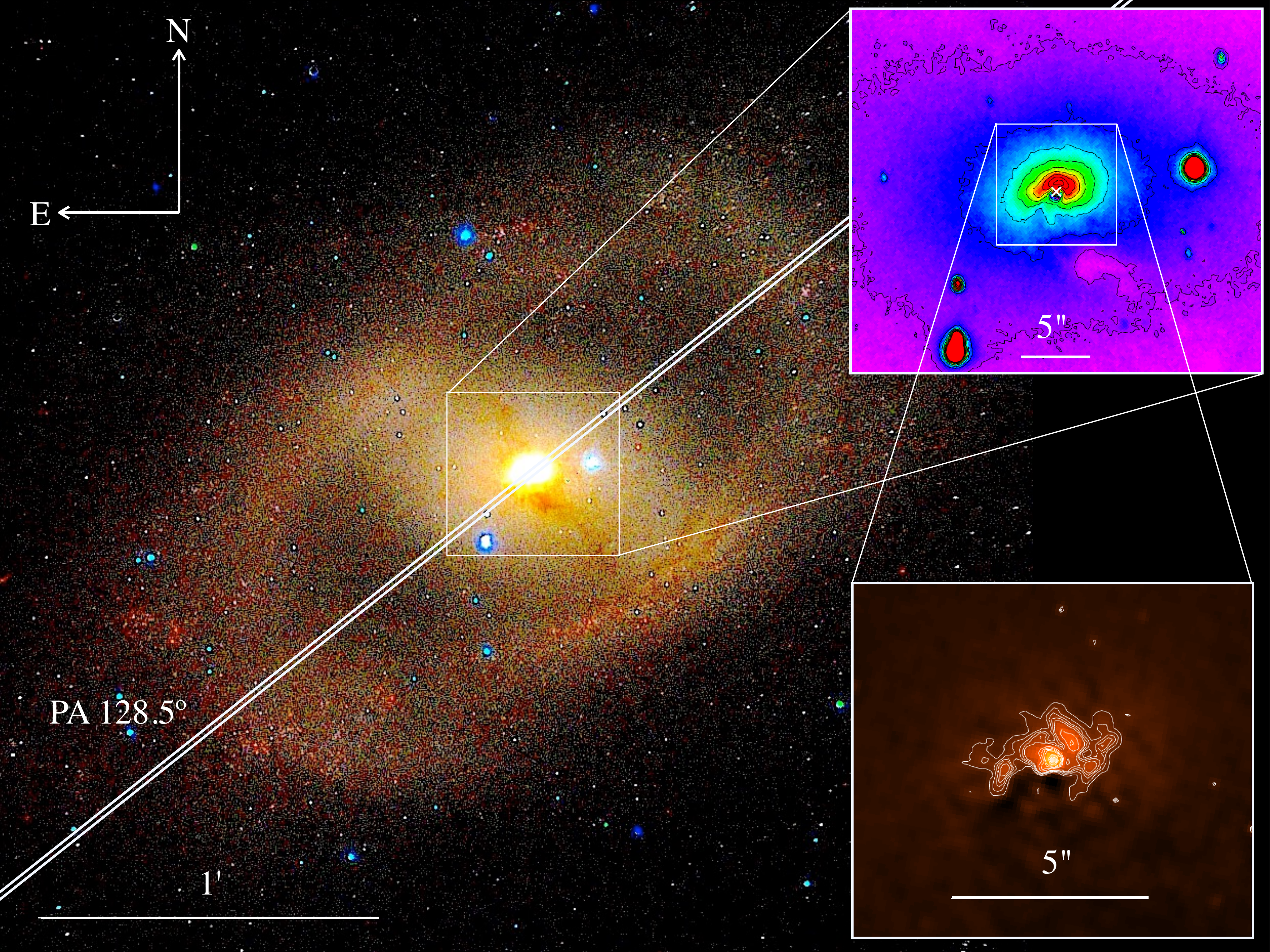}
  \caption{Main panel: $JHK_S$ image composition taken with F2. The slit position used for the spectroscopy is shown at PA $128.5\degree$. The slit is $4\arcmin .4$ long so the ends are outside the plot. The slit width of $0\arcsec .54$ is not at scale in the figure. Up right panel:  $J-K_s$ color map (F2). Bottom right panel: unsharp mask image HST-F814W. }
  \label{Kimage}
\end{figure*}

In the main panel of Figure \ref{Kimage} a $J$, $H$ and $K_s$ RGB composition is shown. This false color image allows us to appreciate the whole extension of the stellar ring that surrounds the galaxy. 

The total apparent magnitude of the galaxy in $K_s$ band up to the 20\,arcsec$^{-2}$ isophote resulted in m$_{Ks} \sim6.8\pm0.4$, while the central object in the nucleus, inside 0\farcs45, reached m$_{Ks}\sim11.7\pm0.4$. This is consistent with the magnitudes reported by \cite{Peng2006}, who performed a component decomposition with GALFIT on 2MASS images and found a disk magnitude of m$_{Ks} \sim 6.15\pm0.5$ and a nuclear magnitude of m$_{Ks} \sim 11.22\pm0.2$. The slight differences between both pairs of magnitudes could have various sources. On the one hand, the higher resolution of Flamingos-2 allows us to determine smaller seeing limited apertures to measure the nuclear magnitude, thus giving a larger value. On the other hand, differences in the zero-point determined while performing the photometry could cause a systematic difference between the magnitudes measured by Peng et al. and ours.

The $JHK_s$ composed image presented in Figure \ref{Kimage} allows us to appreciate the galaxy substructures: the bar is clearly identified and dust lanes are prominent, causing great obscuration to the circumnuclear region of the galaxy and along the bar. Each dust lane is appreciated with different intensity. The northern band probably lies below the bar or is simply more extincted due to the inclination of the galaxy. It is remarkable how both dust lanes cross the bar almost directly from the ring to the circumnuclear region, as is usually observed in weak bars, without delineating the edges of the bar \citep{Martini2004}. The projected length of the bar is $(1.00\pm0.04)\arcmin$, and the angle between the bar and the major axis of the galaxy is $82\degree$. The major axis of an ellipse containing the bulge has an angle of $27\degree$ with the bar major axis. This projection difference between the bulge and the disk would indicate that the bulge has a triaxial shape. We also detect ansae structures at the tip of the bar, as brightness enhancements in the inner ring.

Three different-scale views of NGC\,6300 are depicted in Figure \ref{Kimage}. The main panel, as mentioned above, shows the global scale from F2 $JHK_s$ composed image. In the top right panel, the $J-K_s$ color map constructed from F2 images, allows us to confirm the great extinction present in the south of the nucleus, assumed to be the intensity peak of the F2 $K_s$-band image. In the color map, it can be seen how the southern dust lane reaches the very nucleus, indicating that this lane is probably the cause of the nuclear extinction and the one responsible for the color asymmetry that stands out from the map. This high extinction is not present along the direction of the slit that we used for the spectroscopy. In that direction (P.A. =$128\degree$) no asymmetries are visible. 

In the right bottom panel an unsharp masked image made from the F814W $HST$\footnote{Instrument: WFC3/UVIS, Obs ID: ib6w62060, Exp. Time: 5922.003 sec., Proposal PI: O'Connell, Proposal ID:11360.}  image is presented, depicting the inner 5\arcsec\, of the nuclear region. White contours delineate the central isophotes. No circumnuclear ring can be seen in this image, either it does not exist or lies below the HST spatial resolution. Furthermore, although three arms can be identified, the structure is probably better represented by a disc and chaotic obscuration than by a nuclear spiral, as we will see below in the photometric profile fitting.   

\begin{figure*}[!t] 
  \centering
   \includegraphics[width=1\textwidth]{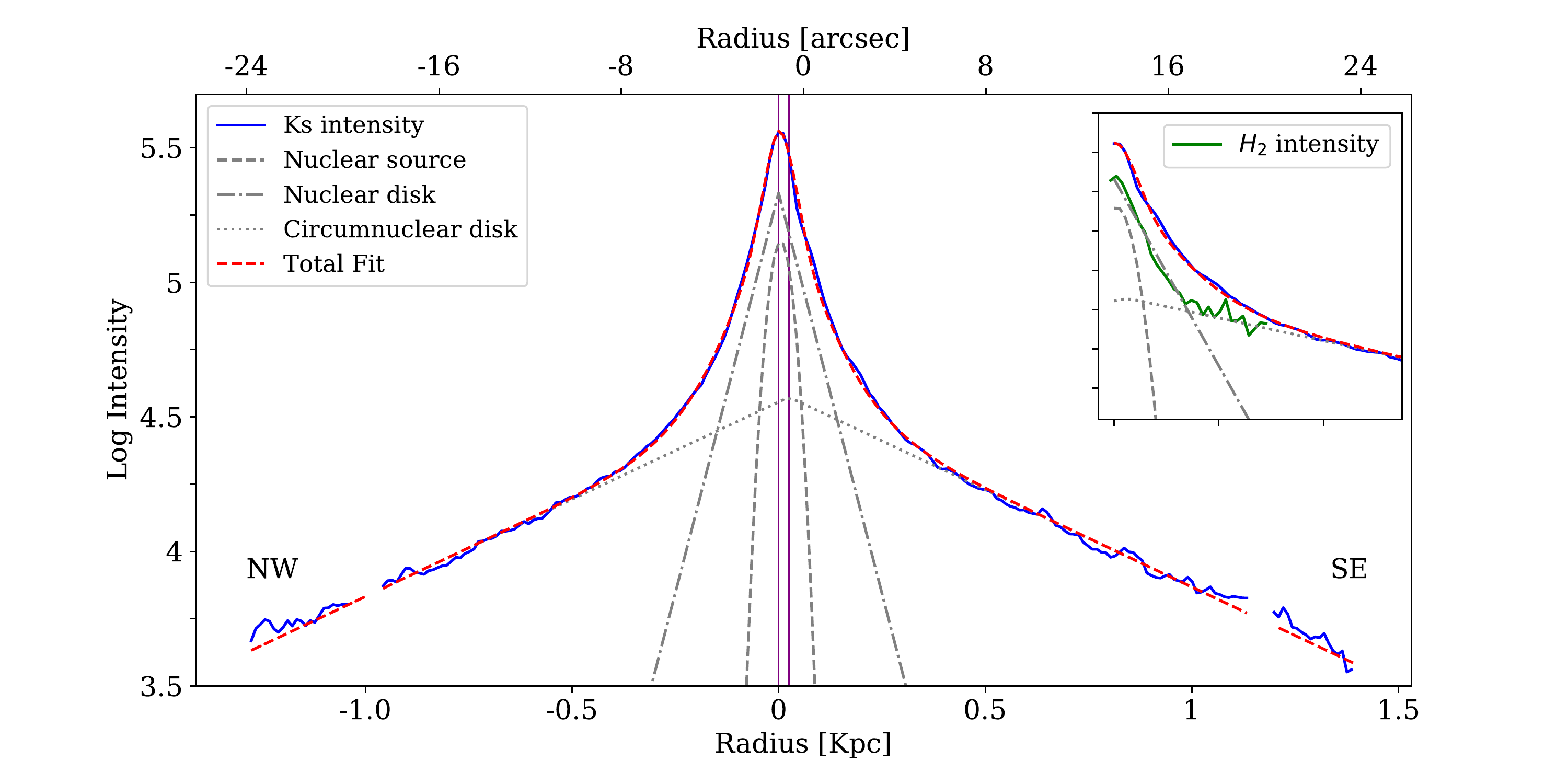}
 \caption{ Main panel: Spatial profile along $PA=128.5^{o}$ extracted from the $K_s$-band image (blue). The different structural components are plotted in gray. The total fit, in red, corresponds to the sum of three components. Two of them correspond to disk structures, one nuclear with a scale radius of 0.073 $\pm$ 0.005\,kpc and a circumnuclear one with a scale radius of 0.60 $\pm$ 0.03\,kpc. Finally, the nuclear source is well fitted by a Gaussian function with an FWHM = 0.07 $\pm$ 0.01\,kpc. The nuclear disk and the central source compose a central structure that is displaced in 25\,pc with respect to the circumnuclear disk (marked with vertical lines). Upper right panel: Zoom-in of the nuclear region for positive radii superimposed to the $H_2  2.12  \mu m$ emission profile which coincides with the nuclear disc component.}
  \label{perfiles}
\end{figure*}

We have performed a 5 pixel (0\farcs9) width extraction on the $K_s$ image, corresponding to 1.5 seeing disk. The radial emission profile obtained (Figure \ref{perfiles}) was fitted with several components in order to model the nuclear structure of this galaxy. In this figure, we show the best fit which consists of a very bright Gaussian component representing the central source with FWHM = 0.07 $\pm$ 0.01\,kpc, a nuclear disk with a scale radius of 0.073 $\pm$ 0.005\,kpc, and finally, a low brightness circumnuclear disk with scale radius of 0.60 $\pm$ 0.03\,kpc. The nuclear structure (Gaussian + nuclear disk) dominates up to 0.25\,kpc and is displaced 25\,pc in the NW direction with respect to the circumnuclear disk. This off-centering will result of particular relevance in Sec. \ref{velcurves} when we analyze the circular velocity curve.
The upper right panel of Figure \ref{perfiles} is a zoom-in of the nuclear region radial profile with the $H_2 2.12 \mu m$ intensity profile superimposed. The $H_2$ profile was extracted from the $K_s$ band 2D spectrum and is well fitted with the same nuclear disk component as the $K_s$-band total profile.       

\subsection{Spectral Analysis}\label{spectra}

At NGC\,6300 distance, and considering F2 pixel scale (0\farcs18 pixel$^{-1}$), one pixel corresponds to 11\,pc and the physical spatial resolution is 33\,pc. For each spectral band we performed seven extractions: one extraction centered in the continuum peak, and three extractions at each side of the peak, centered at $\pm0\farcs44, \pm0\farcs9,\,\, $and$\, \pm 1\farcs44$, which correspond to physical radii of $\pm27$ pc, $\pm55$ pc, and $\pm88$ pc respectively. Positive radius was considered to the southeast direction. The width of each extraction was variable. To determine these, both the spatial resolution limited by seeing (0\farcs6) and the independent information that could be obtained from each spectrum were taken into account.

In order to analyze the nuclear emission with the least amount of contamination by off-nuclear emission regions, the continuum peak extraction was performed with 1 pixel width (0\farcs18) without compromising the S/N.

\begin{figure*}[!t]
  \centering
  \includegraphics[width=1\textwidth]{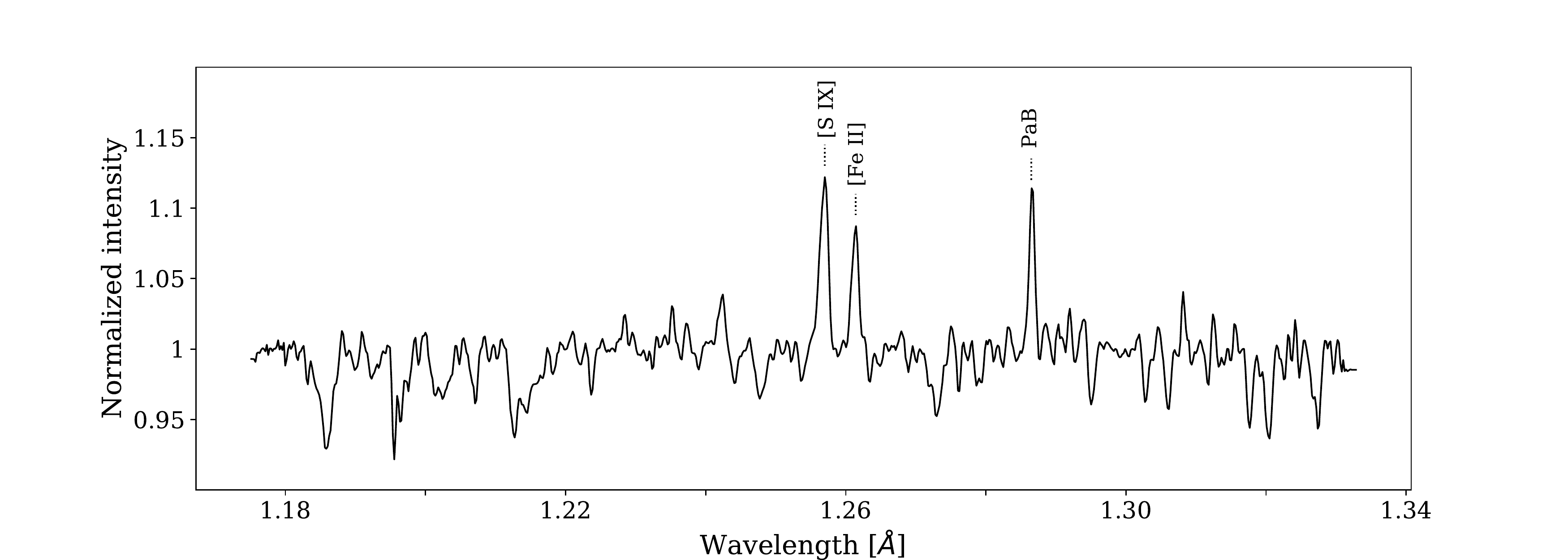}
    \caption{Nuclear $J$ band spectrum of NGC\,6300, the extraction is 1 pixel wide equivalent to 11\,pc. Highlighted are the  [S \textsc{ix}] $\lambda 1.25 \mu m$ coronal line, the [Fe\,\textsc{ii}] $\lambda 1.26 \mu m$ low ionization line, and the the Pa$\beta$ $\lambda 1.28 \mu m$ hydrogen recombination line.}
  \label{Jespectro}
\end{figure*}

The $J$ band nuclear spectrum is shown in Figure \ref{Jespectro}. It shows three prominent emission lines corresponding to, from left to right, [S \textsc{ix}] $\lambda 1.25 \mu m$, [Fe \textsc{ii}]$\lambda 1.26 \mu m$, and the  Pa$\beta$ $\lambda 1.28\mu m$ recombination line.

The presence of the coronal line [S\,\textsc{ix}] $\lambda 1.25 \mu m$ is indicative of extreme UV and X-ray photons, because a very hard spectrum is necessary to reach the high ionization potential of 328.2\,eV of this line (see, for example, \citealt{Reunanen2003}, \citealt{Rodriguez-Ardila2011}, for a detailed treatment of coronal lines). This line is a clear indicator of nuclear activity as nothing but matter falling into an accretion disk can produce such an energetic continuum.

\begin{figure*}[!t] 
  \centering
  \includegraphics[width=0.8\textwidth]{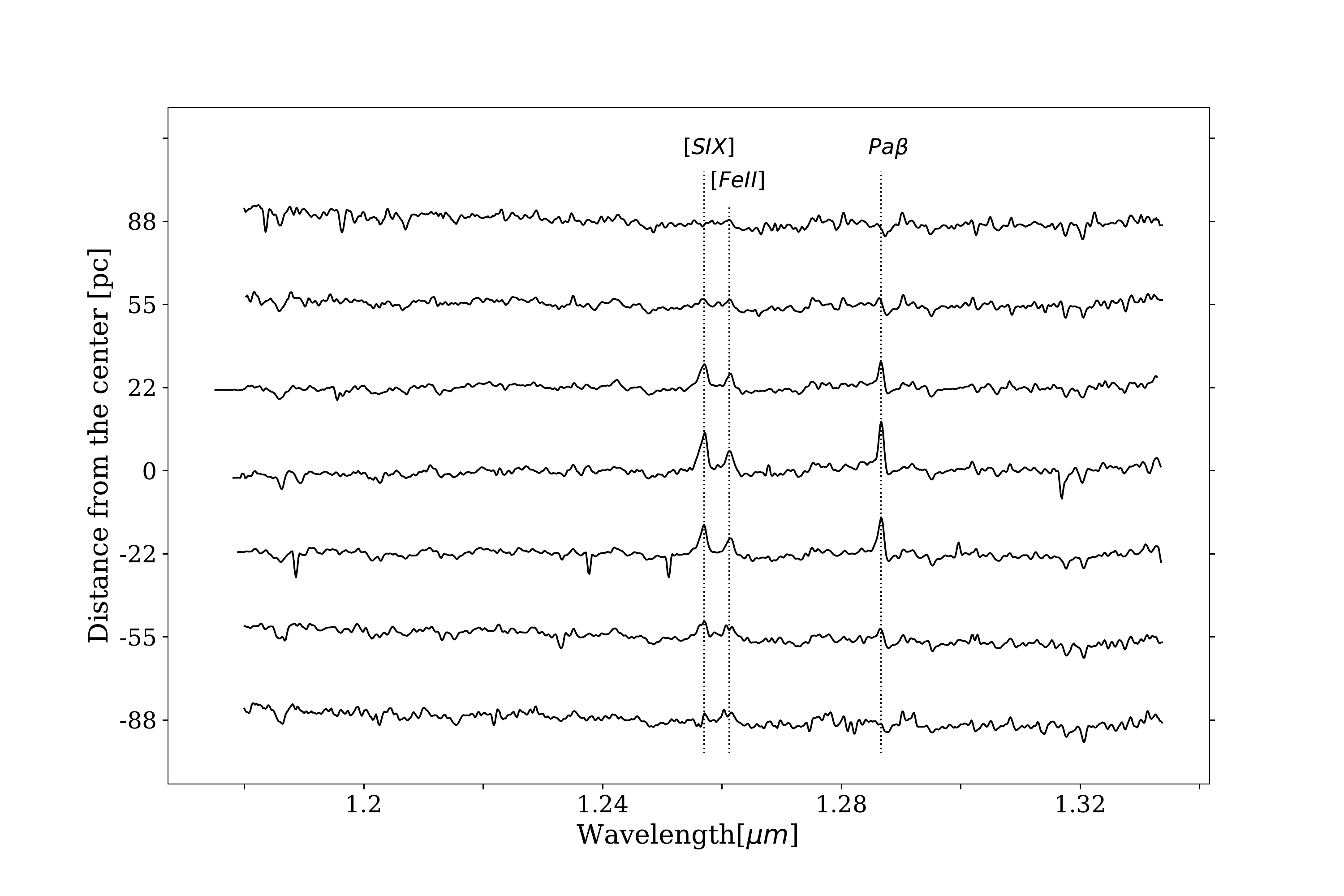}
    \caption{$J$-band spectra for different radii. Highlighted are the [S \textsc{ix}] $\lambda 1.25 \mu m$ coronal line, the [Fe \textsc{ii}] $\lambda 1.26 \mu m$ low ionization line and the Pa$\beta$ $\lambda 1.28 \mu m$ hydrogen recombination line.}
  \label{Jstack}
\end{figure*}

As can be seen from the extractions performed to the $J$-band spectrum (Figure \ref{Jstack}), we detect [S\,\textsc{ix}] emission up to 88 pc in the SE direction, far into the NLR \citep{Padovani2017}. At first sight, this coronal line shows a blueshifted component probably arising from an outflow superimposed to the typical NLR emission component. The $J$ spectra also show the low ionization line [Fe \textsc{ii}] $\lambda 1.26 \mu m$ (IP=7.9 eV) coexisting with the coronal gas for all radii up to 88\,pc in both directions. Pa$\beta$ is present in emission throughout the inner 110 pc, and also in absorption, due to the underlying stellar population. A detailed treatment of the line profiles is presented in Section \ref{lineprofile}.

\begin{figure*}[!t]
  \centering
  \includegraphics[width=1\textwidth]{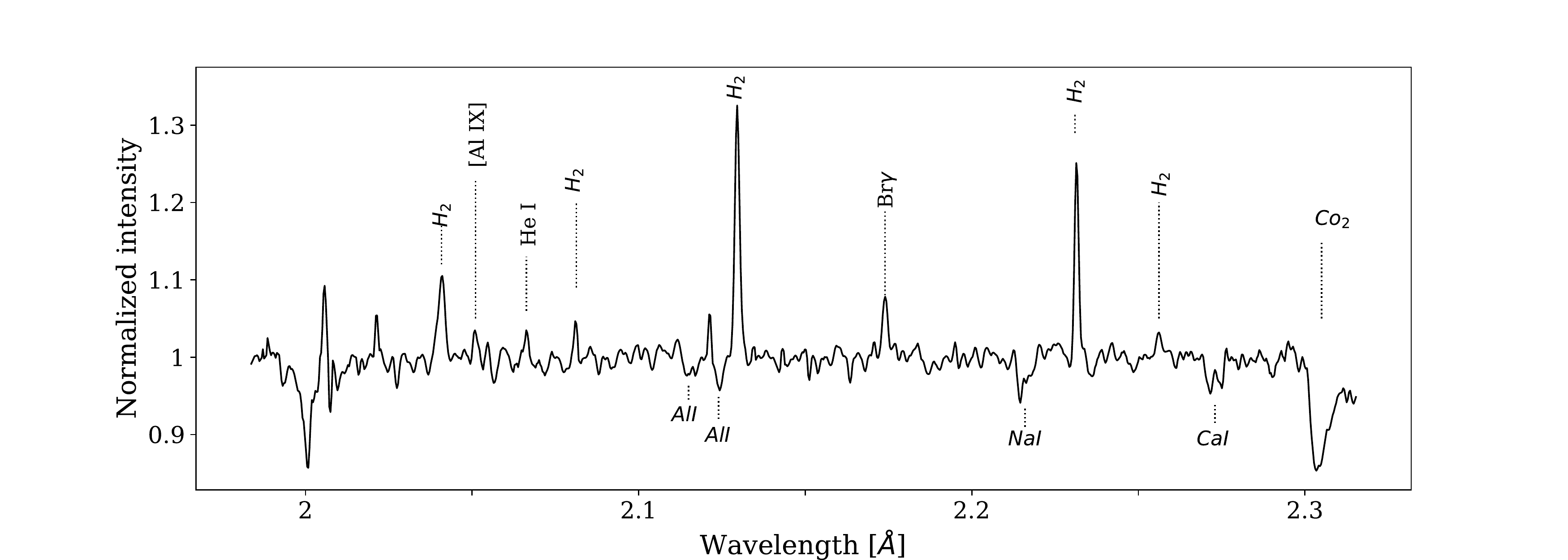}
    \caption{Nuclear $K_s$-band spectrum of NGC\,6300, the extraction is 1 pixel wide, equivalent to 11\,pc. Highlighted are the [Al \textsc{ix}] $\lambda 2.04 \mu m$ high ionization line, the [He \textsc{i}] $\lambda 2.06 \mu m$ line, the Br$\gamma$ $\lambda 2.16 \mu m$ hydrogen recombination line and several $H_{2}$ rotational and vibrational lines. Also present are various absorption lines and doublets produced in the galaxy stellar population.}
  \label{Kespectro}
\end{figure*}

The nuclear extraction of the $K_s$-band spectra is presented in Figure \ref{Kespectro}. The most prominent emission lines are from molecular hydrogen $H_{2}\lambda 2.03\mu$m, $H_{2}\lambda 2.07\mu$m,  $H_{2}\lambda 2.12\mu$m, $H_{2}\lambda 2.22\mu$m, $H_{2}\lambda 2.25\mu$m. Also, the forbidden coronal aluminum line [Al \textsc{ix}] is present at $\lambda 2.04\mu$m, and the Br$\gamma$ hydrogen recombination line at $\lambda 2.16\mu m$. Several absorption lines are visible, including the Na\,\textsc{i} and Ca\,\textsc{i} doublets, and the beginning of the CO$_2$ molecular band.

\begin{figure*}[!t]
  \centering
 \includegraphics[width=0.8\textwidth]{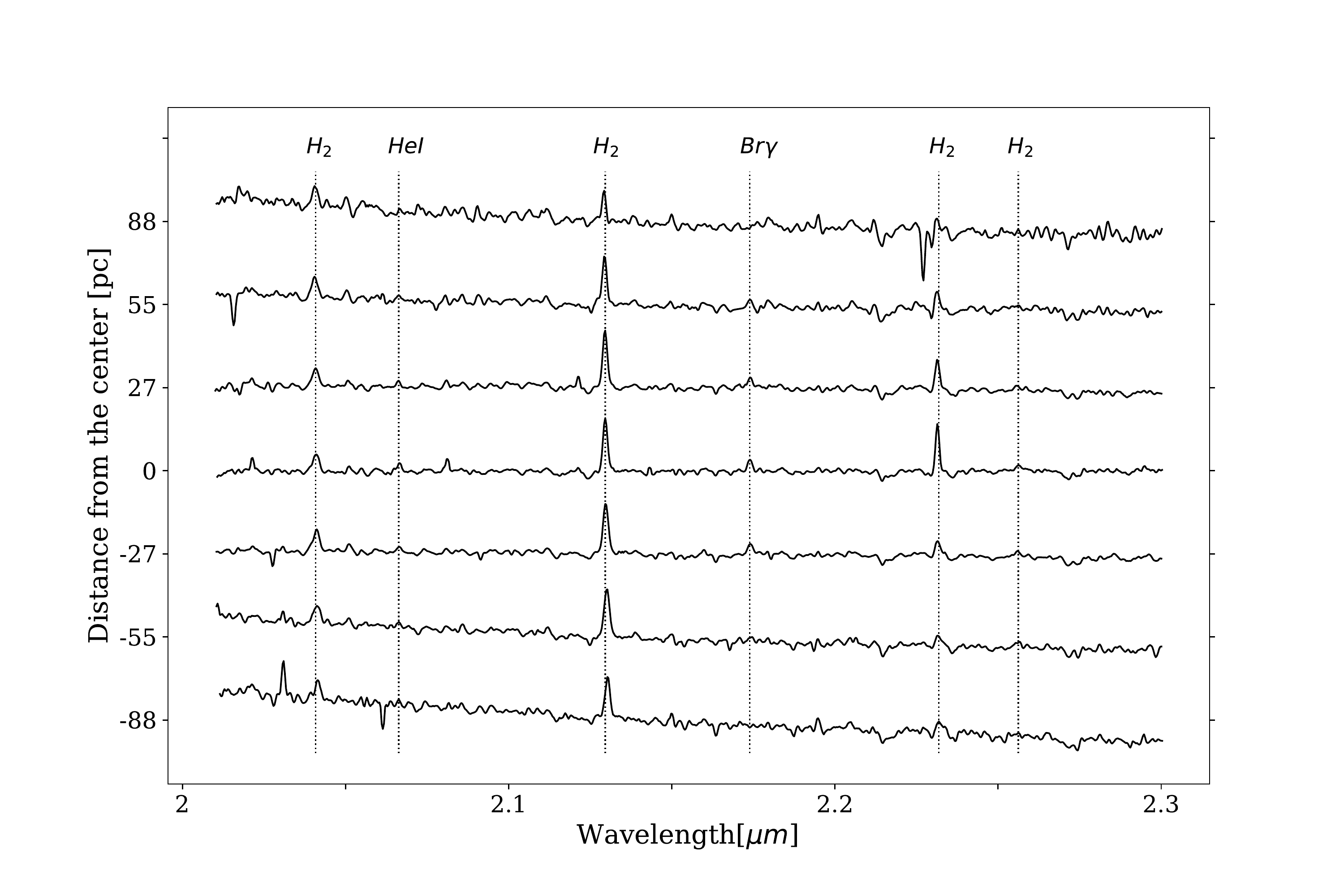}
  \caption{$K_s$-band spectra for different radii. All the extractions are 3 pixels wide in accordance with the seeing spatial resolution limitation, except the innermost extraction which is 1 pixel wide. The comparison between the spectra shows that at the very nucleus the continuum is quite flat and becomes steeper with increasing radius. The main spectral features are highlighted.}
  \label{$K_s$tack}
\end{figure*}

Figure \ref{$K_s$tack} shows all the $K_s$-band extractions for different radii. As for the Pa$\beta$ line, Br$\gamma$ emission is detected until 55 pc, but the $H_{2}$ molecular emission lines are present in all the extractions, so we can infer molecular gas properties up to 88 pc. In Section \ref{H2} we will give more details on these molecular lines and the gas properties.

We used the $IRAF$ task \textit{splot} to fit Gaussian functions to all the emission lines in the $J$- and $K_s$-band spectra to determine the intensity, flux, $FWHM$, and radial velocity shift of each line. In the following sections we present the results obtained for the molecular gas kinematics, the properties derived from the line ratios, and the dust properties emerging from the analysis of the continuum.

\subsubsection{The $H_{2}$ excitation mechanism}\label{H2} 

The $K_s$ band presents a richness of hydrogen molecular lines that can be used to test the main excitation mechanism of this element. Pioneering work was made theoretically and observationally by several authors in the 1980 and 1990 decades (see for example \citealt{Tanaka89} and \citealt{Mouri1994} and references therein). In these works it is established that $H_2$ has two main excitation mechanisms that often happen simultaneously: (1) thermal excitation caused either by shocks, UV radiation or X-ray radiation, and (2) non-thermal excitation associated with UV fluorescence. It is possible to distinguish between these mechanisms because they occur associated with different vibrational and rotational temperatures that produce different line intensity ratios. In the first case, the vibrational and rotational temperatures are essentially equal and therefore this state can be characterized by a single excitation temperature typically in the range 1000-2000\,K. In the non-thermal case, the vibrational temperature is much higher than the rotational, occurring this conditions  in colder gas.

Following \cite{Reunanen2002}, the vibrational and rotational temperatures can be calculated from the $H_{2}$ $1 - 0S(2) \lambda 2.03 \mu m /1 -0S(0) \lambda 2.22\mu $m  and $1 - 0S(1)\lambda  2.12 \mu m /2 - 1S(1)\lambda  2.25 \mu $m line ratios as follows:

\begin{equation}
T_{vib} \simeq 5600/ln(1.355 x I_{1-0S(1)}/I_{2-1S(1)}
\end{equation}

\begin{equation}
T_{rot} \simeq -1113/ln(0.323 x I_{1-0S(2)}/I_{1-0S(0)}
\end{equation}

These line ratios have also the advantage of being little affected by extinction due to the proximity of the lines and of the effect of the ortho/para ratio of the molecules because 2-1S(1) and 1-0S(1) occur in ortho-molecules while 1-0S(2) and 1-0S(0) in para molecules \citep{Mouri1994}.

\begin{figure}[!t]
  \centering
   \includegraphics[width=0.5\textwidth]{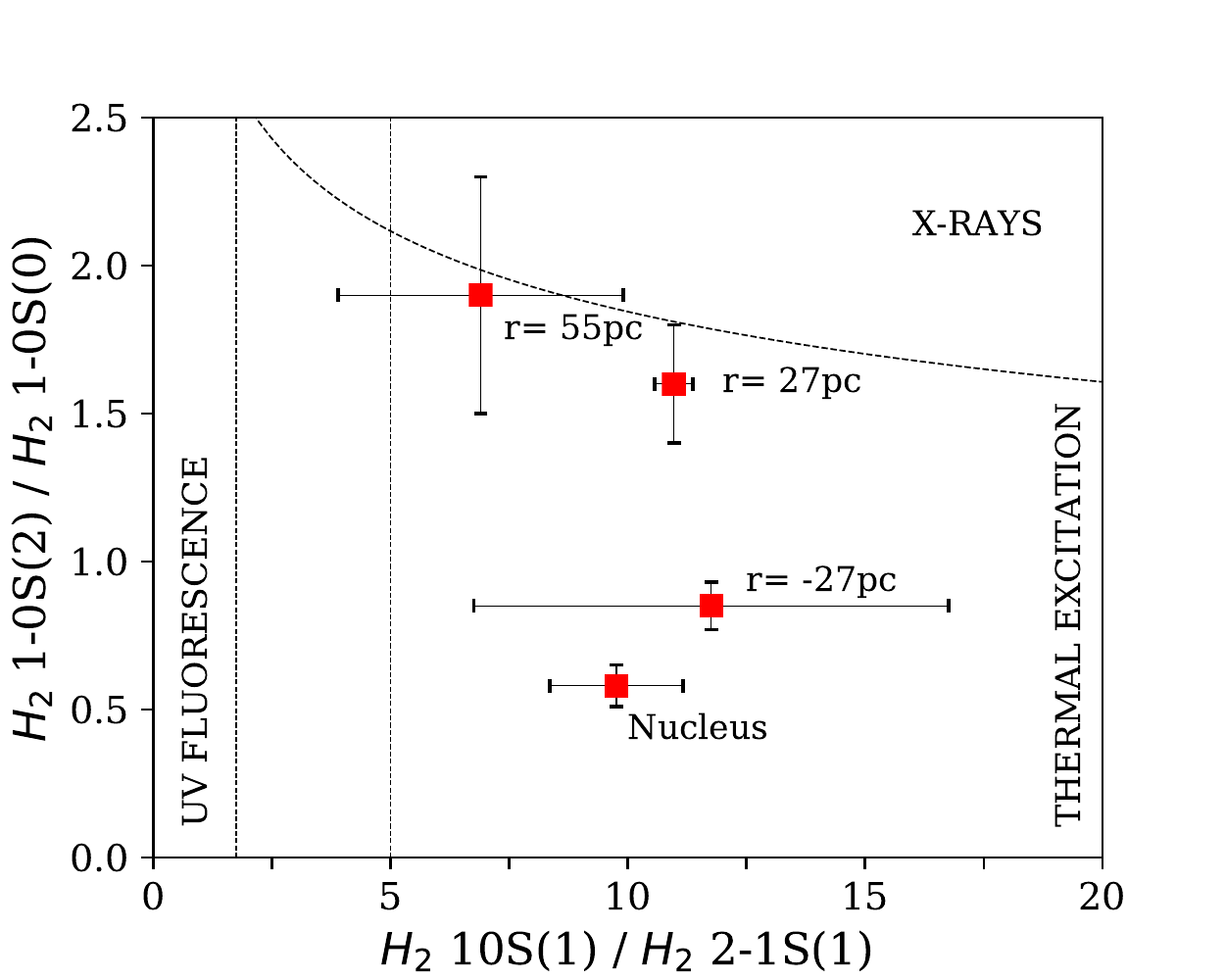}
 \caption{Data points corresponding to the $H_{2}$ $1 - 0S(1)/2 - 1S(1)$ vs   $1 - 0S(2) /1 -0S(0)$ line intensity ratios. The radius of extraction is indicated. The dashed lines separate regions in the diagram where different excitation mechanisms dominate. The upper curve corresponds to $T_{viv}=T_{rot}$ and the vertical lines separates the thermal excitation region from the UV excitation region. In the intermediate region a mixture of excitation mechanisms occurs. }
  \label{diagramaH2}
\end{figure}

The dominant excitation mechanism can be determined from these line ratios as can be seen in Figure \ref{diagramaH2} (\citealt{RamosAlmeida2009}, \citealt{FalconBarroso2014}, \citealt{Gunthardt2015}). The value of $T_{vib} \leqslant 6000$  separates the thermal domain from the nonthermal while the top right corner curve corresponds to $T_{rot} = T_{viv}$ and above the curve lies the region where the excitation through X-ray photons dominates. Figure \ref{diagramaH2} represents a diagnostic diagram for NGC\,6300 calculated for the different radii. Our data points lie in the thermal region where either shocks or collisional processes prevail. 

\subsubsection{$H_{2}$ radial velocity curve and black hole mass} \label{velcurves}

\begin{figure*}[!t] 
  \centering
   \includegraphics[trim=1cm 5.5cm 2cm 5.5cm,clip=true,width=\textwidth]{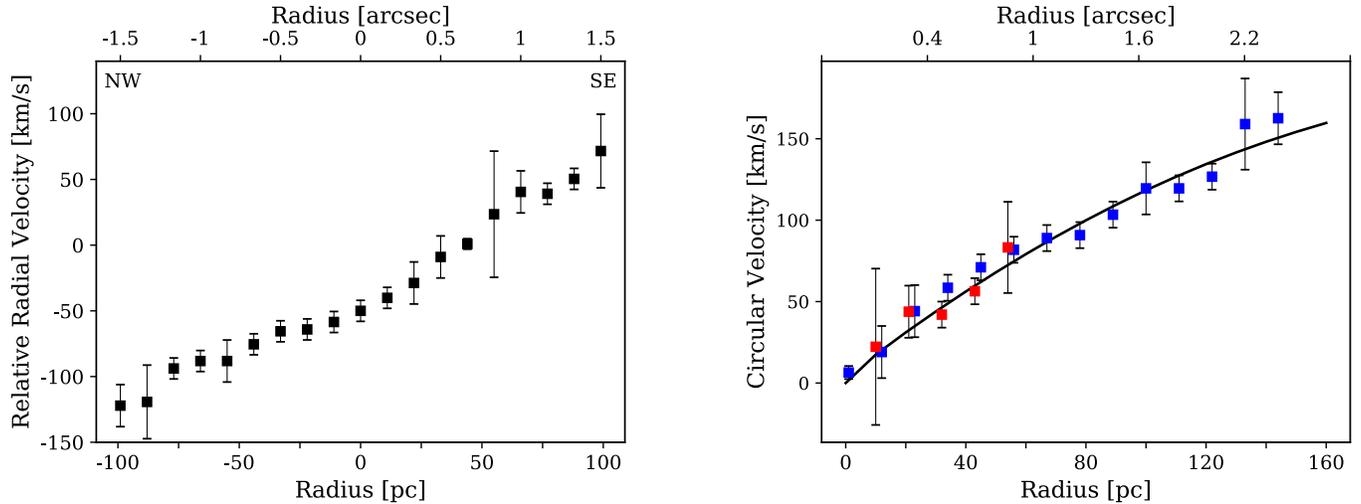}
 \caption{\textit{Left:} Radial velocity curve derived from the  $\lambda 2.12 \mu$m molecular hydrogen emission line. The data points correspond to the velocity of the line at each radius relative to the nucleus. \textit{Right:} Circular velocity curve from the same H$_2$ where the inclination was assumed to be $i=(52\pm1)\degree$. The blue points correspond to reflected negative radii. The fit is a composition of a Plummer plus a Hernquist mass models. The origin of the plot is placed in the true kinematic center (see text for details).}
  \label{vr}
 \end{figure*}

We derived the molecular gas radial velocity curve by measuring the velocity shift of the strongest emission line $H_{2}\lambda 2.12\mu$m for each radius as mentioned above. We also measured the $H_{2}$ radial velocities directly from the 2D spectrum as an alternate method obtaining similar results. This allowed us to quantify the uncertainties and to be confident with our error estimations. The radial velocity curve is presented in Figure \ref{vr} (\textit{left}) were it is clearly seen that the gas follows a rotation pattern in the innermost 200 pc. Error bars were determined performing several measurements of each line position and taking the standard deviation.

The circular velocity curve presented in Figure \ref{vr} (\textit{right}) was constructed by deprojecting the measured velocities by the galaxy inclination $i=(52\pm1)\degree$ and reflecting the negative velocities towards positive radii. When negative velocities are reflected, it is observed that the curve is not symmetric about the continuum peak value. A symmetric curve could be obtained by displacing the curve 55 pc in the SW direction. The blue points in the curve correspond to the negative values of the radial velocity curve in Figure \ref{vr} (\textit{left}), once displaced 55\,pc in radius and, therefore, $\sim50$ km s$^{-1}$.  This new radius value that symmetrizes the radial velocity curve, would be the true rotation center of the galaxy (according to our PA), and is not coincident with the $K_{s}$-band continuum peak. This displacement of the emission peak with respect to the rotation center was already detected by \cite{Buta1987} from optical spectra. \cite{Buta1987} raises the possibility of both an asymmetric mass distribution, and that the dust is obscuring the true nucleus of the galaxy. The author is more inclined towards the second option, however he addresses the need to obtain high-resolution NIR spectra to unravel this issue. Our observations allow us to determine this off-centering with precision, both from the kinematics, and from the $K_s$-band brightness profile as we have shown in Section \ref{structure}. Reinforcing the idea of a true mass distribution asymmetry. Figure \ref{vr} (\textit{right}) also shows a two component model fitting with a Plummer model of $4\times10^9\,\Msun$ and scale radius of 240\,pc, and a Hernquist model of $1.8\times10^{10}\,\Msun$ and scale radius of 4400\,pc. In Sec. \ref{structure} we have presented an off-centering of 25\,pc  according to our surface brightness profile fitting in Figure \ref{perfiles}, between the inner structure, conformed by the nuclear source plus a nuclear disk and a circumnuclear disk that begins to dominate from $\sim0.25$ kpc outwards. This off-centering, determined from the photometric structural components, together with the displacement of the rotation center obtained from the circular velocity curve, allow us to infer that the global symmetry center, the peak of the continuum in $K_{s}$ band, and the dynamic rotation center of the galaxy, would not be coincident. In this scenario, the SMBH would be at the $K_{s}$ band continuum peak but in orbit around the true rotation center of the galaxy as derived from the circular velocity curve.

Could this asymmetry be responsible for the chaotic distribution of dust in the inner 300\,pc? And furthermore, could this be at least part of the elusive mechanism that produces the angular momentum loss in the inner pc needed to feed the AGN? These off-centerings have been found both in active and inactive galaxies such as NGC\,3227 \citep{Arribas1994}, NGC\,1068 \citep{Arribas1996}, NGC\,1672 \citep{Diaz99}, M\,83 \citep{Mast06}, and NGC\,253 \citep{Gunthardt2015} among others, and predicted by simulations \citep{Emsellem2015}.

Finally, from the radial velocity curve we can infer that the molecular gas is confined into a rotating disk as seen also in the $H_2$ intensity profile (Fig. \ref{perfiles}). The absence of turbulent or chaotic motion as well as no broad component in the emission line profiles associated with outflows, indicate that this molecular disk should be shielded from the AGN ionizing continuum, probably by large amounts of dust. 
This scenario is reinforced by the emission line diagnostic diagram presented in Figure \ref{diagramaH2} which shows that the main excitation mechanism of the molecular hydrogen is photoionization or shocks, probably produced by young stars and supernovae and no by X-rays originated in the accretion disk of the black hole.

With the gas velocity measurements in the central 5 points (inner 55\,pc) we were able to estimate the enclosed mass. In a Keplerian approximation and taking into account the inclination of the galaxy, the mass enclosed in a radius R [pc]  which presents a velocity amplitude V [km s$^{-1}$] is:

\begin{equation}
M = 233\, V^2R/sen(i)
\end{equation}

Where the resulting mass is in solar masses.
We used the IRAF \textit{ellipse} task for determining the inner inclination performing an ellipse fitting to the F2 $K_s$ band image. Our fit justifies the use of an inclination $i=(52\pm1)\degree$ for the mass determination in accordance with previous works \citep{Vacu1991} where the inclination was determined from coarser data. If we assume, by angular momentum conservation, that the innermost part of the circumnuclear disk has the same inclination, we can estimate an enclosed mass as an upper limit for the dark object residing in the very nucleus.

With the assumptions mentioned above, we derived:

\begin{equation}
(M_{SMBH}/{\Msun})^{upper}=(6.2\pm 2.5) \times 10^{7}
\end{equation}

for the material inside 55 pc, and:

\begin{equation}
(M_{SMBH}/{\Msun})^{upper}=(6.9\pm 5.1) \times 10^{6} 
\end{equation}

for the mass enclosed in the minimum resolved diameter of 33 pc.

This is in accordance with previous estimations from different methods (\citealt{Beifiori2009}, \citealt{Khor2012},  \citealt{Davis2014}) and would place the mass of the central compact object into the intermediate-to-low black hole mass regime.

\subsubsection{The $K_s$ band continuum slope}\label{kcont}

The continuum of the $K_s$ band spectra presents an interesting shape, which evidences the radial variation of the properties of the different sources that contribute to it, in particular, the stellar population and the circumnuclear dust. From Figure \ref{$K_s$tack} we can see that the nuclear spectrum is quite flat while for increasing radii the continuum slope becomes steeper. This flattening has been interpreted as arising from the composed contributions of the red tail of the stellar continuous emission and the blue tail of dust heated by the AGN for which emission peaks in the midinfrared (e.g. \citealt{Edelson1986}, \citealt{Alonso-Herrero1998}, \citealt{Ferruit2004}, \citealt{RamosAlmeida2009}, \citealt{Prieto2010}). This continuum slope behavior is ubiquitous in type 1 nuclei (e.g. \citealt{Kobayashi1993}), and have been observed in some type 2 nuclei. For example, \cite{Martins2010} found a very steep $K_s$-band continuum in NGC\,1068 and suggest that it is caused by thermal radiation from hot dust.
In other work, \cite{Ferruit2004} find a similar behavior of the continuum $K_s$-band slope for NGC\,2110 as we do in Figure \ref{$K_s$tack}. Their extractions are performed each $1\arcsec$ which represents 172\,pc at NGC\,2110 distance. Probably due to lower spatial resolution, the authors found this infrared excess only in the nuclear spectrum. 

In general, works in the literature present a spatial sampling that does not allow them to resolve the circumnuclear region below 100\,pc. In this way the infrared excess is only detectable in the central spectrum. This has led them to assume that the origin of this excess is the dusty torus that surrounds the AGN, in accordance with the standard unified model.

Our Infrared data provide a spatial sampling of 33\,pc, which allow us to resolve the radial circumnuclear variations of the $K_s$-band continuum slope in detail, and find that the flattening caused by dust is not only observed in the nuclear spectrum but also in the circumnuclear regions, at least up to 27\,pc. This means that the presence of warm dust is not exclusive of the nuclear region, but also of more extended regions.

\begin{figure*}[!t]
  \centering
\includegraphics[width=1.1\textwidth]{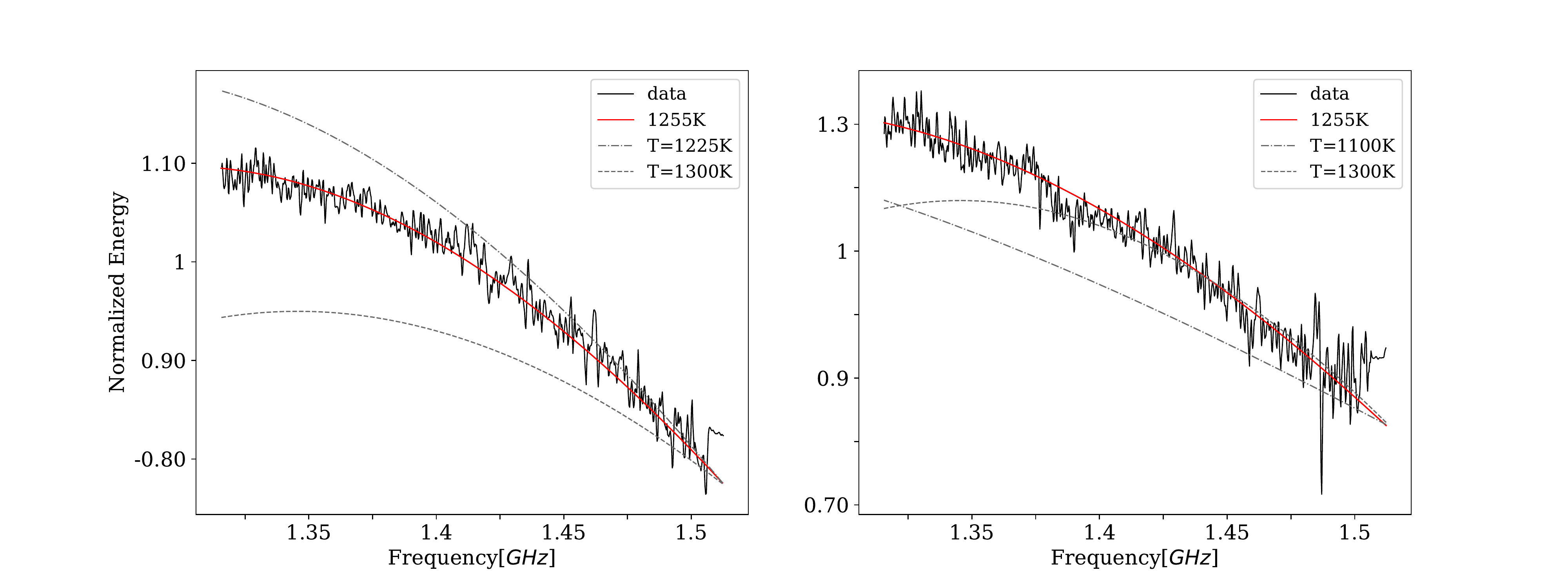}
 \caption{Red excess obtained after subtracting a mean external spectrum from a central one. Left panel: the nuclear spectrum. Right panel: the spectrum corresponding to -27\,pc. In both cases the red curve represents the best single temperature fit and the gray curves are examples of near temperatures for comparison.  }
  \label{polvo}
\end{figure*}

In order to test if a warm dust component producing a red excess is responsible for the flattening in the nuclear spectra, we proceed as follows. First, we construct a pseudo-stellar population template averaging the two external spectra in  $\pm88$\,pc. Assuming that the stellar population does not vary significantly in the small region considered, we subtracted this template from the inner spectrum and from the immediate adjacent spectrum at 27\,pc. The results of these subtractions, shown in Figure \ref{polvo}, once converted into Frequency vs. Normalized Energy, resembles a thermal emission component rising through lower frequencies. We can fit a family of blackbody spectra to the infrared excess and compare the slopes to constrain the temperature of the warm dust. Both spectra can be fitted by a combination of two or more blackbody spectra but the simpler fitting is in both cases a single temperature, 1255\,K in the nuclear case and 1225\,K in the 27\,pc spectrum. Both temperatures are consistent with emission from hot dust grains (see \citealt{Landt2011} and references therein). Although the real scenario must be more complex, with a continuum of dust temperatures, this fitting points out that the hot component must be present. We have tested that no combination of blackbody below 1200\,K can reproduce the $K_s$-band bump around 2.15 $\mu m$ or equivalently 1.4 GHz. The gray curves in Figure \ref{polvo} correspond to blackbody emission of different temperatures and are only for comparison. It is worth noting that because our spectra are not flux calibrated, only the slopes can be analyzed and the intensities are arbitrary.

In order to reach this temperature, dust molecules must be directly irradiated by hard photons from the AGN, therefore the only possible place for their location is the edge of the ionization cone, since otherwise they would be blown away by the outflow.
Our results point to consider the role of extended dust in the circumnuclear region as a mechanism that participates in the obscuration of the central engine, without contradicting the existence of the dusty torus itself, confined into the central parsec of the UM. This supports the scenario where the whole dust distribution in the hundreds of central parsecs plays an important role in the AGN classification in addition to the inclination of the nuclear region.

\subsubsection{Ionized gas line profile}\label{lineprofile}

\begin{figure}[!t]
  \centering
\includegraphics[width=0.5\textwidth]{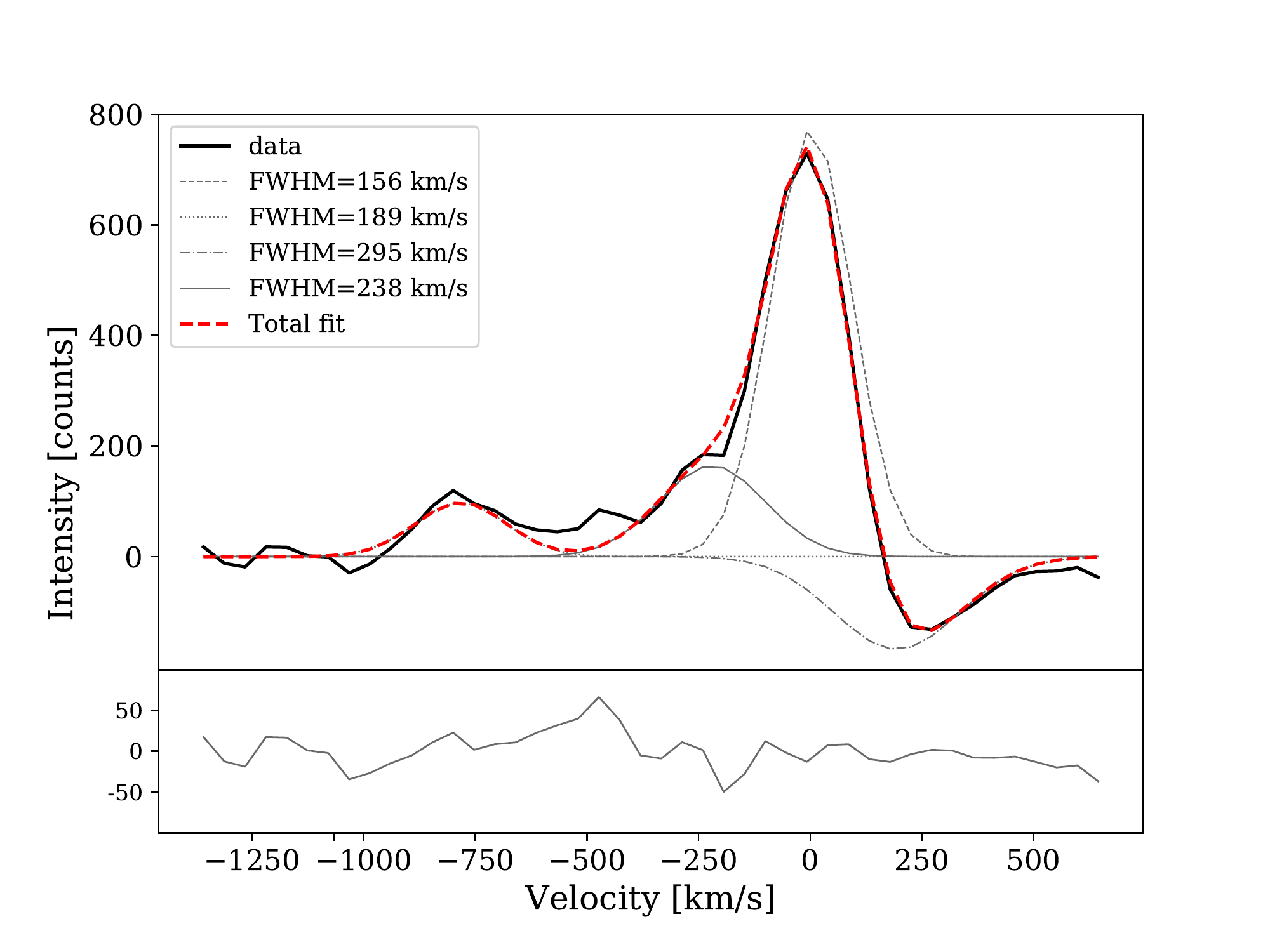}
\includegraphics[width=0.5\textwidth]{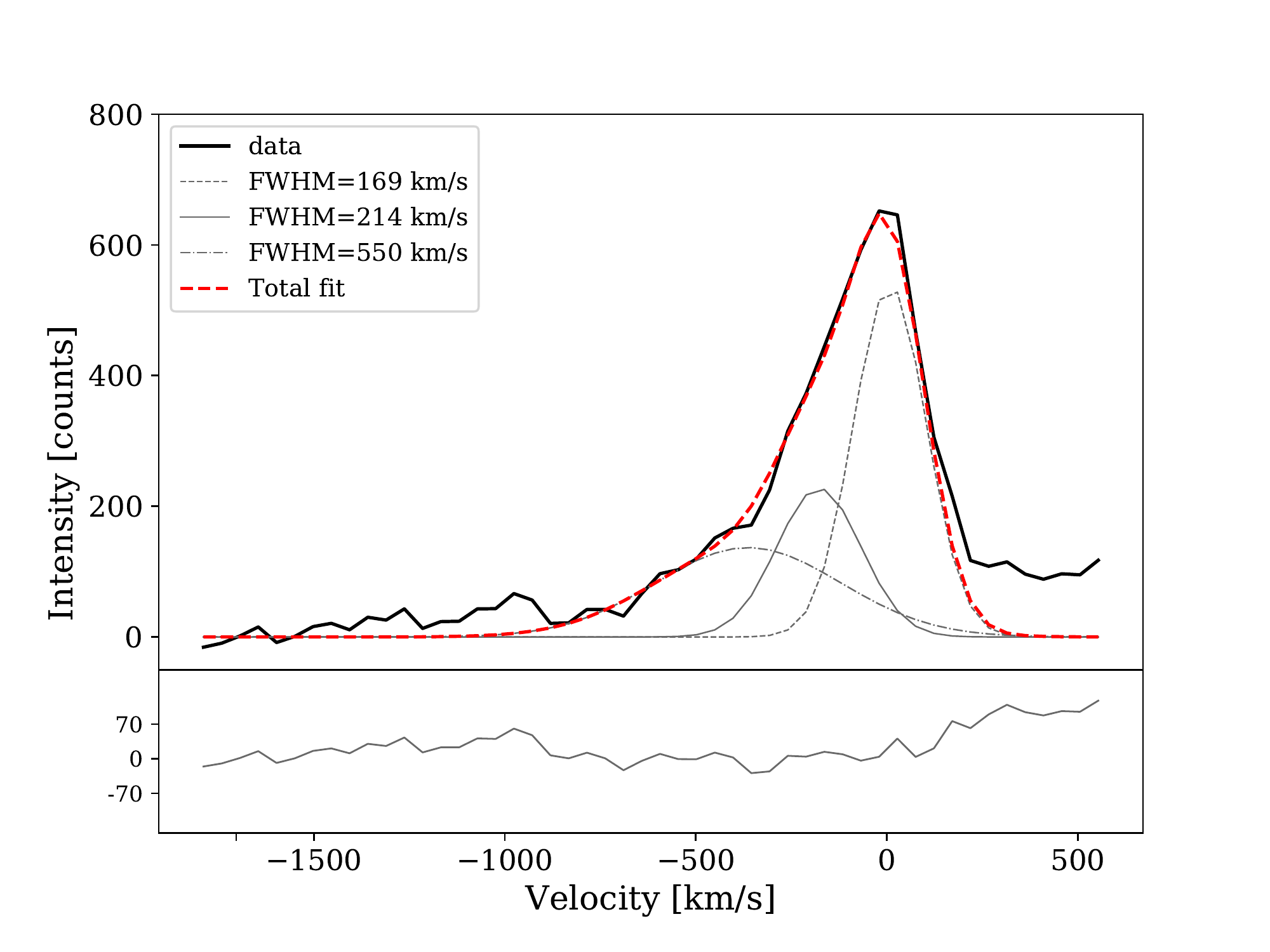}
 \caption{Emission line component fittings. In black, the data extracted from the nuclear spectrum of 33 pc wide is indicated in black, the different components are in gray, and the total fit to the line is in red. Top: The Pa$\beta$ $\lambda 1.28\mu$m line. The negative component accounts for the absorption in the underlying stellar population of NGC\,6300. Bottom: The [S\,\textsc{ix}] $\lambda 1.25 \mu m$ line, the red excess in the residual corresponds to the blue tail of the [Fe \textsc{ii}]$\lambda 1.26 \mu m$ line. Note that the first two components (dashed and solid components) are the same, within the errors, in both lines, suggesting cospatiality of the emitting regions.}
  \label{PaB}
\end{figure}

Analyzing the line profiles for the $J$ band, a blueshifted component is observed in all lines and at all radii where the lines are observable. Figure \ref{PaB} shows the fit for Pa$\beta \lambda 1.28\mu$m and [S\,\textsc{ix}]$\lambda 1.25 \mu m$ lines. In both cases, a single gaussian is far from a good fit, so more than one component is mandatory. For Pa$\beta$ the contribution of the underlying stellar population must be taken into account. The [Fe\,\textsc{ii}]$\lambda 1.26 \mu m$  line cannot be modeled because it is too close to the [S\,\textsc{ix}] line and presents lower S/N, hence its profile is severely affected. The red part of the [S\,\textsc{ix}] profile is also affected, but, because no stellar absorption is expected for this coronal line, the only consequence of the blending is a small red tail that do not affect the fit significantly.

The [S\,\textsc{ix}] sulfur line  shows an asymmetry towards the blue more pronounced than the Pa$\beta$ profile. If we assume for the [S\,\textsc{ix}] profile fitting the same widths as for the main narrow and broad Pa$\beta$ components, the need for a third component is inferred from the residual. This third component is not visible in the Pa$\beta$ profile, we attribute this to the fact that the spectrum is noisier in this region than around $\lambda 1.25 \mu m$ and that the part of the spectrum redward Pa$\beta$ is affected by the telluric correction up to $\sim \lambda 1.27 \mu m$.

For the Pa$\beta$ line, the fit consists in a narrow $(156\pm10)$ km s$^{-1}$ width typical of type-2 AGN main component \citep{Osterbrock2006}, a $(295 \pm 50)$ km s$^{-1}$ negative stellar absorption component redshifted in $(191\pm 20)$ km s$^{-1}$ (consistent with the other stellar absorption lines present in the spectra), a broader $(238\pm 20)$ km s$^{-1}$ component blueshifted in $(223\pm3)$ km s$^{-1}$, and a $(189\pm 10)$ km s$^{-1}$ component blueshifted in $(783\pm 10)$ km s$^{-1}$ that is strongly affected by noise.

For the [S \textsc{ix}] line, the fit consists in a $(169\pm20)$ km s$^{-1}$ width typical of type-2 AGN main component \citep{Oster1989}, a broader $(214\pm20)$ km s$^{-1}$ component blueshifted in $(186 \pm 5)$ km s$^{-1}$ and a broad $(550\pm40)$ km s$^{-1}$ component blueshifted in $(371 \pm 30)$ km s$^{-1}$.

The  broad blueshifted components present in these lines are  probably associated with outflow processes. Furthermore, while the ionizing source of the Pa$\beta$ emission could be a mixing of star formation and AGN radiation,  [S\,\textsc{ix}] could only be ionized by hard photons from the accretion disc, as mentioned before Furthermore, because the main components of both lines present the same kinematic width, within the errors, they must be cospatial,  meaning that the star formation contribution, if present, is not observable with this combination of spectral resolution and S/N. Both the emission and the outflow components in Pa$\beta$ and [S\,\textsc{ix}] are detected up to a distance of 55 pc from the nucleus.

The co-existence of coronal lines in the spectrum, which shows that we are seeing the NLR, and emission lines with lower ionization potential, indicates that these NLR clouds are thick enough to shield the region that opposes the AGN from its hard ionizing radiation.

\section{Summary and conclusions} \label{summary}

In this paper we have studied the nuclear region of NGC\,6300 analyzing $J, K_s$, and $H$ images and $J$- and $K_s$-band spectra acquired with the Gemini-South Flamingos-2 imaging spectrograph. To achieve this goal, we analyzed the spatial profile in $K_s$-band images from which we could discern the different structural components that make up the central kiloparsecs. We have also performed spectral extractions to study the radial variations of the spectroscopic properties of both dust and ionized gas, in the inner 500\,pc. Our main conclusions can be summarized as follows:

\begin{itemize}
  \setlength\itemsep{-0.3em}
\item{We performed a fitting to the $K_s$-band photometric profile and found that the structure that dominates the central 0.25\,kpc is composed of a central Gaussian source with FWHM = 0.07 $\pm$ 0.01\,kpc and a nuclear disk with scale radius 0.073 $\pm$ 0.005\,kpc. This structure presents an off-centering of 25\,pc with respect to the circumnuclear disk that dominates up to the kiloparsec scale and presents a scale radius of 0.60 $\pm$ 0.03\,kpc .}

\item{From the $H_{2}$ circular velocity curve we measured an off-centering of 55\,pc between the kinematical center and emission peak associated with the SMBH.}
\item{We have detected in the $J$-band spectra [S \textsc{ix}] $\lambda 1.25 \mu m$, [Fe \textsc{ii}]$\lambda 1.26 \mu m$, and Pa$\beta$ $\lambda 1.28\mu m$. The presence of the coronal line [S \textsc{ix}] $\lambda 1.25 \mu m$ up to 88\,pc from the nucleus is indicative of extreme UV and X-ray photons.}
\item{[S \textsc{ix}] and Pa$\beta$  both present a blue-shifted broad components we associate with nuclear outflows.}
\item{In the $K_s$-band spectra we have identified molecular hydrogen $H_{2}\lambda 2.03\mu$m, $H_{2}\lambda 2.07\mu$m,  $H_{2}\lambda 2.12\mu$m, $H_{2}\lambda 2.22\mu$m, $H_{2}\lambda 2.25\mu$m. Also a forbidden, coronal Aluminum line at $\lambda 2.04\mu$m and Br$\gamma$ emission at $\lambda 2.16\mu m$. Several absorption lines are visible, including the NaI and CaI doublets, and the CO$_2$ molecular band. According to our diagnostic diagram constructed from line ratios, the $H_{2}$ emission lines are thermally excited.}
\item{From the radial velocity curve constructed from $H_{2}$ emission, the thermal excitation and the absence of outflows, we have confirmed that the molecular hydrogen gas is confined to a rotating disk, shielded from the AGN.}
\item{From a Keplerian approximation we estimated the mass enclosed in the inner $\sim55$\,pc to be $M=(6.2\pm 2.5) \times 10^{7}\,\Msun$, that can be considered as an upper limit for the SMBH mass powering the AGN.}
\item{An analysis of the $K_s$-band continuum slope shows an IR excess outside the nucleus and up to~$\sim$ 27\,pc probably associated with hot dust and not related to the putative dusty torus of the UM. We propose that this hot dust must be contributing to the AGN obscuration and it could be partially responsible for its Sy2 classification.} 

\end{itemize}

Thanks to the spatial resolution reached by Gemini and Flamingos-2, we are able to study the circumnuclear region of NGC\,6300, a highly obscured Sy2 galaxy, at scales of tens of parsecs. Both F2 and $HST$ images allowed us to verify that the nucleus of the galaxy, assumed to be the emission peak in $K_s$-band and where the SMBH resides, is displaced from the geometric center of the circumnuclear disk. The fact that the radial velocity curve is not steep shows that the central gravitational potential is not dominated by the point source that powers the AGN. This SMBH is probably orbiting the kinematical center of the galaxy while accreting mass and growing. This motion may be responsible for the chaotic distribution of gas and dust that is observed in the images of the central hundreds of parsecs. And it is perhaps this wobbling mechanism that causes the angular momentum dissipation of the material forcing it to travel the last parsecs before being accreted onto the SMBH. Moreover, the absence of a circumnuclear ring and strong circularization of orbits that would stop the fall of the material towards the center is in agreement with this lack of gravitational preponderance of the SMBH. The future mass growth of the dark object will increasing the innermost angular speeds, lead to the appearance of Lindblad resonances and the probable formation of the mentioned ring. This grow of the SMBH will also cause the destruction of the weak bar in NGC\,6300, as predicted by numerical simulations. 

The results presented in this work provide different elements that help to understand the way in which an obscured AGN of intermediate-low mass grows its mass in the presence of a weak bar and with large amounts of circumnuclear dust. But there are many questions that arise and require a systematic study in a statistically significant sample. What is the relationship between nuclear feeding mechanisms in Seyfert galaxies and the strength of the bar? How the appearance of the circumnuclear ring, the growth of the SMBH, the AGN duty cycle, and the life cycle of the bar all linked? And finally, how does the distribution of circumnuclear dust influence the Seyfert1/Seyfert2 classification?

All this places obscured AGN as the most appropriate objects to deepen those aspects of the Unified Model that are still under strong debate.

\begin{acknowledgments}
Based on observations obtained at the Gemini Observatory, which is operated by the Association of Universities for Research in Astronomy, Inc., under a cooperative agreement with the NSF on behalf of the Gemini partnership: the National Science Foundation (United States), the National Research Council (Canada), CONICYT (Chile), Ministerio de Ciencia, Tecnolog\'{i}a e Innovaci\'{o}n Productiva (Argentina), and Minist\'{e}rio da Ci\^{e}ncia, Tecnologia e Inova\c{c}\~{a}o (Brazil). 
This research has made use of the NASA/IPAC Extragalactic Database (NED), which is operated by the Jet Propulsion Laboratory, California Institute of Technology, under contract with the National Aeronautics and Space Administration. Based on observations made with the NASA/ESA Hubble Space Telescope, obtained from the data archive at the Space Telescope Science Institute. STScI is operated by the Association of Universities for Research in Astronomy, Inc. under NASA contract NAS 5-26555. G.G. has a fellowship from Consejo Nacional de 
Investigaciones Cient\'ificas y T\'ecnicas, CONICET, Argentina. This paper has been partially supported with grants from CONICET and Secretar\'ia de Ciencia y Tecnolog\'ia, Universidad Nacional de C\'ordoba, SeCyT-UNC, Argentina. We want to thank Horacio Dottori and Percy Gomez for fruitful discussions, and Bruno S\'anchez for shearing his python expertise. We thank an anonymous referee for many helpful comments that improved the quality of this paper.  

\end{acknowledgments}


\bibliographystyle{yahapj}
\bibliography{references}

\begin{thebibliography}{}
\providecommand\natexlab[1]{#1}
\providecommand\JournalTitle[1]{#1}

\bibitem[{{Alonso-Herrero} {et~al.}(1998){Alonso-Herrero}, {Simpson}, {Ward},
  \& {Wilson}}]{Alonso-Herrero1998}
{Alonso-Herrero}, A., {Simpson}, C., {Ward}, M.~J., \& {Wilson}, A.~S. 1998,
  \href{http://dx.doi.org/10.1086/305269}{\JournalTitle{\apj}, 495, 196}

\bibitem[{{Alonso-Herrero} {et~al.}(2011){Alonso-Herrero}, {Ramos Almeida},
  {Mason}, {Asensio Ramos}, {Roche}, {Levenson}, {Elitzur}, {Packham},
  {Rodr{\'{\i}}guez Espinosa}, {Young}, {D{\'{\i}}az-Santos}, \&
  {P{\'e}rez-Garc{\'{\i}}a}}]{AlonsoHerrero2011}
{Alonso-Herrero}, A., {Ramos Almeida}, C., {Mason}, R., {et~al.} 2011,
  \href{http://dx.doi.org/10.1088/0004-637X/736/2/82}{\JournalTitle{\apj}, 736,
  82}

\bibitem[{{Antonucci}(1993)}]{Antonucci1993}
{Antonucci}, R. 1993,
  \href{http://dx.doi.org/10.1146/annurev.aa.31.090193.002353}{\JournalTitle{\araa},
  31, 473}

\bibitem[{{Arribas} \& {Mediavilla}(1994)}]{Arribas1994}
{Arribas}, S., \& {Mediavilla}, E. 1994,
  \href{http://dx.doi.org/10.1086/174983}{\JournalTitle{\apj}, 437, 149}

\bibitem[{{Arribas} {et~al.}(1996){Arribas}, {Mediavilla}, \&
  {Garcia-Lorenzo}}]{Arribas1996}
{Arribas}, S., {Mediavilla}, E., \& {Garcia-Lorenzo}, B. 1996,
  \href{http://dx.doi.org/10.1086/177265}{\JournalTitle{\apj}, 463, 509}

\bibitem[{{Awaki} {et~al.}(2005){Awaki}, {Murakami}, {Leighly}, {Matsumoto},
  {Hayashida}, \& {Grupe}}]{Awaki2005}
{Awaki}, H., {Murakami}, H., {Leighly}, K.~M., {et~al.} 2005,
  \href{http://dx.doi.org/10.1086/433161}{\JournalTitle{\apj}, 632, 793}

\bibitem[{{Beckmann} {et~al.}(2006){Beckmann}, {Gehrels}, {Shrader}, \&
  {Soldi}}]{Beckmann2006}
{Beckmann}, V., {Gehrels}, N., {Shrader}, C.~R., \& {Soldi}, S. 2006,
  \href{http://dx.doi.org/10.1086/499034}{\JournalTitle{\apj}, 638, 642}

\bibitem[{{Beifiori} {et~al.}(2009){Beifiori}, {Sarzi}, {Corsini}, {Dalla
  Bont{\`a}}, {Pizzella}, {Coccato}, \& {Bertola}}]{Beifiori2009}
{Beifiori}, A., {Sarzi}, M., {Corsini}, E.~M., {et~al.} 2009,
  \href{http://dx.doi.org/10.1088/0004-637X/692/1/856}{\JournalTitle{\apj},
  692, 856}

\bibitem[{{Bianchi} {et~al.}(2005){Bianchi}, {Guainazzi}, {Matt}, {Chiaberge},
  {Iwasawa}, {Fiore}, \& {Maiolino}}]{Bianchi2005}
{Bianchi}, S., {Guainazzi}, M., {Matt}, G., {et~al.} 2005,
  \href{http://dx.doi.org/10.1051/0004-6361:20053389}{\JournalTitle{\aap}, 442,
  185}

\bibitem[{{Blandford} \& {Rees}(1978)}]{Blandford1978}
{Blandford}, R.~D., \& {Rees}, M.~J. 1978,
  \href{http://dx.doi.org/10.1088/0031-8949/17/3/020}{\JournalTitle{\physscr},
  17, 265}

\bibitem[{{Buta}(1987)}]{Buta1987}
{Buta}, R. 1987, \href{http://dx.doi.org/10.1086/191199}{\JournalTitle{\apjs},
  64, 383}

\bibitem[{{Davis} {et~al.}(2014){Davis}, {Berrier}, {Johns}, {Shields},
  {Hartley}, {Kennefick}, {Kennefick}, {Seigar}, \& {Lacy}}]{Davis2014}
{Davis}, B.~L., {Berrier}, J.~C., {Johns}, L., {et~al.} 2014,
  \href{http://dx.doi.org/10.1088/0004-637X/789/2/124}{\JournalTitle{\apj},
  789, 124}

\bibitem[{{de Vaucouleurs} {et~al.}(1991){de Vaucouleurs}, {de Vaucouleurs},
  {Corwin}, {Buta}, {Paturel}, \& {Fouque}}]{Vacu1991}
{de Vaucouleurs}, G., {de Vaucouleurs}, A., {Corwin}, Jr., H.~G., {et~al.}
  1991, \JournalTitle{\skytel}, 82, 621

\bibitem[{{D{\'{\i}}az} {et~al.}(1999){D{\'{\i}}az}, {Carranza}, {Dottori}, \&
  {Goldes}}]{Diaz99}
{D{\'{\i}}az}, R., {Carranza}, G., {Dottori}, H., \& {Goldes}, G. 1999,
  \href{http://dx.doi.org/10.1086/306781}{\JournalTitle{\apj}, 512, 623}

\bibitem[{{D{\'{\i}}az} {et~al.}(2006){D{\'{\i}}az}, {Dottori}, {Aguero},
  {Mediavilla}, {Rodrigues}, \& {Mast}}]{Diaz2006}
{D{\'{\i}}az}, R.~J., {Dottori}, H., {Aguero}, M.~P., {et~al.} 2006,
  \href{http://dx.doi.org/10.1086/507886}{\JournalTitle{\apj}, 652, 1122}

\bibitem[{{D\'{i}az} {et~al.}(2013){D\'{i}az}, {G\'omez}, {Schirmer},
  {Navarrete}, {Stephens}, {Bosch}, {Gaspar}, {Camperi}, \&
  {Gunthardt}}]{Diaz2013}
{D\'{i}az}, R.~J., {G\'omez}, P., {Schirmer}, M., {et~al.} 2013,
  \JournalTitle{Boletin de la Asociacion Argentina de Astronomia La Plata
  Argentina}, 56, 457

\bibitem[{{Dullemond} \& {van Bemmel}(2005)}]{Dullemond2005}
{Dullemond}, C.~P., \& {van Bemmel}, I.~M. 2005,
  \href{http://dx.doi.org/10.1051/0004-6361:20041763}{\JournalTitle{\aap}, 436,
  47}

\bibitem[{{Edelson} \& {Malkan}(1986)}]{Edelson1986}
{Edelson}, R.~A., \& {Malkan}, M.~A. 1986,
  \href{http://dx.doi.org/10.1086/164479}{\JournalTitle{\apj}, 308, 59}

\bibitem[{{Eikenberry} {et~al.}(2012){Eikenberry}, {Bandyopadhyay}, {Bennett},
  {Bessoff}, {Branch}, {Charcos}, {Corley}, {Dewitt}, {Eriksen}, {Elston},
  {Frommeyer}, {Gonzalez}, {Hanna}, {Herlevich}, {Hon}, {Julian}, {Julian},
  {Lasso}, {Marin-Franch}, {Marti}, {Murphey}, {Raines}, {Rambold}, {Rashkind},
  {Warner}, {Leckie}, {Gardhouse}, {Fletcher}, {Hardy}, {Dunn}, {Wooff}, \&
  {Pazder}}]{Eikenberry2012}
{Eikenberry}, S., {Bandyopadhyay}, R., {Bennett}, J.~G., {et~al.} 2012,
  \href{http://dx.doi.org/10.1117/12.925679}{in \procspie, Vol. 8446,
  Ground-based and Airborne Instrumentation for Astronomy IV}, 84460I

\bibitem[{{Elitzur}(2012)}]{Elitzur2012}
{Elitzur}, M. 2012,
  \href{http://dx.doi.org/10.1088/2041-8205/747/2/L33}{\JournalTitle{\apjl},
  747, L33}

\bibitem[{{Elitzur} \& {Shlosman}(2006)}]{Elitzur2006}
{Elitzur}, M., \& {Shlosman}, I. 2006,
  \href{http://dx.doi.org/10.1086/508158}{\JournalTitle{\apjl}, 648, L101}

\bibitem[{{Elvis} {et~al.}(2004){Elvis}, {Risaliti}, {Nicastro}, {Miller},
  {Fiore}, \& {Puccetti}}]{Elvis2004}
{Elvis}, M., {Risaliti}, G., {Nicastro}, F., {et~al.} 2004,
  \href{http://dx.doi.org/10.1086/424380}{\JournalTitle{\apjl}, 615, L25}

\bibitem[{{Emsellem} {et~al.}(2015){Emsellem}, {Renaud}, {Bournaud},
  {Elmegreen}, {Combes}, \& {Gabor}}]{Emsellem2015}
{Emsellem}, E., {Renaud}, F., {Bournaud}, F., {et~al.} 2015,
  \href{http://dx.doi.org/10.1093/mnras/stu2209}{\JournalTitle{\mnras}, 446,
  2468}

\bibitem[{{Fabian}(2012)}]{Fabian2012}
{Fabian}, A.~C. 2012,
  \href{http://dx.doi.org/10.1146/annurev-astro-081811-125521}{\JournalTitle{\araa},
  50, 455}

\bibitem[{{Falc{\'o}n-Barroso} {et~al.}(2014){Falc{\'o}n-Barroso}, {Ramos
  Almeida}, {B{\"o}ker}, {Schinnerer}, {Knapen}, {Lan{\c c}on}, \&
  {Ryder}}]{FalconBarroso2014}
{Falc{\'o}n-Barroso}, J., {Ramos Almeida}, C., {B{\"o}ker}, T., {et~al.} 2014,
  \href{http://dx.doi.org/10.1093/mnras/stt2189}{\JournalTitle{\mnras}, 438,
  329}

\bibitem[{{Ferruit} {et~al.}(2004){Ferruit}, {Mundell}, {Nagar}, {Emsellem},
  {P{\'e}contal}, {Wilson}, \& {Schinnerer}}]{Ferruit2004}
{Ferruit}, P., {Mundell}, C.~G., {Nagar}, N.~M., {et~al.} 2004,
  \href{http://dx.doi.org/10.1111/j.1365-2966.2004.08009.x}{\JournalTitle{\mnras},
  352, 1180}

\bibitem[{{Goulding} \& {Alexander}(2009)}]{Goulding2009}
{Goulding}, A.~D., \& {Alexander}, D.~M. 2009,
  \href{http://dx.doi.org/10.1111/j.1365-2966.2009.15194.x}{\JournalTitle{\mnras},
  398, 1165}

\bibitem[{{Goulding} {et~al.}(2012){Goulding}, {Alexander}, {Bauer}, {Forman},
  {Hickox}, {Jones}, {Mullaney}, \& {Trichas}}]{Goulding2012}
{Goulding}, A.~D., {Alexander}, D.~M., {Bauer}, F.~E., {et~al.} 2012,
  \href{http://dx.doi.org/10.1088/0004-637X/755/1/5}{\JournalTitle{\apj}, 755,
  5}

\bibitem[{{Greusard} {et~al.}(2000){Greusard}, {Friedli}, {Wozniak},
  {Martinet}, \& {Martin}}]{Greusard2000}
{Greusard}, D., {Friedli}, D., {Wozniak}, H., {Martinet}, L., \& {Martin}, P.
  2000, \href{http://dx.doi.org/10.1051/aas:2000250}{\JournalTitle{\aaps}, 145,
  425}

\bibitem[{{Guainazzi}(2002)}]{Guainazzi2002}
{Guainazzi}, M. 2002,
  \href{http://dx.doi.org/10.1046/j.1365-8711.2002.05132.x}{\JournalTitle{\mnras},
  329, L13}

\bibitem[{{G{\"u}nthardt} {et~al.}(2015){G{\"u}nthardt}, {Ag{\"u}ero},
  {Camperi}, {D{\'{\i}}az}, {Gomez}, {Bosch}, \& {Schirmer}}]{Gunthardt2015}
{G{\"u}nthardt}, G.~I., {Ag{\"u}ero}, M.~P., {Camperi}, J.~A., {et~al.} 2015,
  \href{http://dx.doi.org/10.1088/0004-6256/150/5/139}{\JournalTitle{\aj}, 150,
  139}

\bibitem[{{Khorunzhev} {et~al.}(2012){Khorunzhev}, {Sazonov}, {Burenin}, \&
  {Tkachenko}}]{Khor2012}
{Khorunzhev}, G.~A., {Sazonov}, S.~Y., {Burenin}, R.~A., \& {Tkachenko}, A.~Y.
  2012,
  \href{http://dx.doi.org/10.1134/S1063773712080026}{\JournalTitle{Astronomy
  Letters}, 38, 475}

\bibitem[{{Kobayashi} {et~al.}(1993){Kobayashi}, {Sato}, {Yamashita}, {Shiba},
  \& {Takami}}]{Kobayashi1993}
{Kobayashi}, Y., {Sato}, S., {Yamashita}, T., {Shiba}, H., \& {Takami}, H.
  1993, \href{http://dx.doi.org/10.1086/172261}{\JournalTitle{\apj}, 404, 94}

\bibitem[{{Lacy} {et~al.}(2013){Lacy}, {Ridgway}, {Gates}, {Nielsen}, {Petric},
  {Sajina}, {Urrutia}, {Cox Drews}, {Harrison}, {Seymour}, \&
  {Storrie-Lombardi}}]{Lacy2013}
{Lacy}, M., {Ridgway}, S.~E., {Gates}, E.~L., {et~al.} 2013,
  \href{http://dx.doi.org/10.1088/0067-0049/208/2/24}{\JournalTitle{\apjs},
  208, 24}

\bibitem[{{Landt} {et~al.}(2011){Landt}, {Elvis}, {Ward}, {Bentz}, {Korista},
  \& {Karovska}}]{Landt2011}
{Landt}, H., {Elvis}, M., {Ward}, M.~J., {et~al.} 2011,
  \href{http://dx.doi.org/10.1111/j.1365-2966.2011.18383.x}{\JournalTitle{\mnras},
  414, 218}

\bibitem[{{Leighly} {et~al.}(1999){Leighly}, {Halpern}, {Awaki}, {Cappi},
  {Ueno}, \& {Siebert}}]{Leighly1999}
{Leighly}, K.~M., {Halpern}, J.~P., {Awaki}, H., {et~al.} 1999,
  \href{http://dx.doi.org/10.1086/307649}{\JournalTitle{\apj}, 522, 209}

\bibitem[{{Malkan} {et~al.}(1998){Malkan}, {Gorjian}, \& {Tam}}]{Malkan1998}
{Malkan}, M.~A., {Gorjian}, V., \& {Tam}, R. 1998,
  \href{http://dx.doi.org/10.1086/313110}{\JournalTitle{\apjs}, 117, 25}

\bibitem[{Martini(2004)}]{Martini2004}
Martini, P. 2004, in Penetrating Bars through Masks of Cosmic Dust, ed. D.~L.
  Block, I.~Puerari, K.~C. Freeman, R.~Groess, \& E.~K. Block (Dordrecht:
  Springer Netherlands), 213

\bibitem[{{Martins} {et~al.}(2010){Martins}, {Rodr{\'{\i}}guez-Ardila}, {de
  Souza}, \& {Gruenwald}}]{Martins2010}
{Martins}, L.~P., {Rodr{\'{\i}}guez-Ardila}, A., {de Souza}, R., \&
  {Gruenwald}, R. 2010,
  \href{http://dx.doi.org/10.1111/j.1365-2966.2010.17042.x}{\JournalTitle{\mnras},
  406, 2168}

\bibitem[{{Mast} {et~al.}(2006){Mast}, {D{\'{\i}}az}, \& {Ag{\"u}ero}}]{Mast06}
{Mast}, D., {D{\'{\i}}az}, R.~J., \& {Ag{\"u}ero}, M.~P. 2006,
  \href{http://dx.doi.org/10.1086/499941}{\JournalTitle{\aj}, 131, 1394}

\bibitem[{{Matsumoto} {et~al.}(2004){Matsumoto}, {Nava}, {Maddox}, {Leighly},
  {Grupe}, {Awaki}, \& {Ueno}}]{Matsumoto2004}
{Matsumoto}, C., {Nava}, A., {Maddox}, L.~A., {et~al.} 2004,
  \href{http://dx.doi.org/10.1086/425566}{\JournalTitle{\apj}, 617, 930}

\bibitem[{{Matt}(2000)}]{Matt2000}
{Matt}, G. 2000, \JournalTitle{\aap}, 355, L31

\bibitem[{{Morganti} {et~al.}(1999){Morganti}, {Tsvetanov}, {Gallimore}, \&
  {Allen}}]{Morganti1999}
{Morganti}, R., {Tsvetanov}, Z.~I., {Gallimore}, J., \& {Allen}, M.~G. 1999,
  \href{http://dx.doi.org/10.1051/aas:1999258}{\JournalTitle{\aaps}, 137, 457}

\bibitem[{{Mouri}(1994)}]{Mouri1994}
{Mouri}, H. 1994, \href{http://dx.doi.org/10.1086/174184}{\JournalTitle{\apj},
  427, 777}

\bibitem[{{Mu{\~n}oz Mar{\'{\i}}n} {et~al.}(2007){Mu{\~n}oz Mar{\'{\i}}n},
  {Gonz{\'a}lez Delgado}, {Schmitt}, {Cid Fernandes}, {P{\'e}rez},
  {Storchi-Bergmann}, {Heckman}, \& {Leitherer}}]{Munoz2007}
{Mu{\~n}oz Mar{\'{\i}}n}, V.~M., {Gonz{\'a}lez Delgado}, R.~M., {Schmitt},
  H.~R., {et~al.} 2007,
  \href{http://dx.doi.org/10.1086/519448}{\JournalTitle{\aj}, 134, 648}

\bibitem[{{M{\"u}ller S{\'a}nchez} {et~al.}(2006){M{\"u}ller S{\'a}nchez},
  {Davies}, {Eisenhauer}, {Tacconi}, {Genzel}, \&
  {Sternberg}}]{MullerSanchez2006}
{M{\"u}ller S{\'a}nchez}, F., {Davies}, R.~I., {Eisenhauer}, F., {et~al.} 2006,
  \href{http://dx.doi.org/10.1051/0004-6361:20054387}{\JournalTitle{\aap}, 454,
  481}

\bibitem[{{M{\"u}ller-S{\'a}nchez} {et~al.}(2018){M{\"u}ller-S{\'a}nchez},
  {Hicks}, {Malkan}, {Davies}, {Yu}, {Shaver}, \& {Davis}}]{MullerSanchez2018}
{M{\"u}ller-S{\'a}nchez}, F., {Hicks}, E.~K.~S., {Malkan}, M., {et~al.} 2018,
  \href{http://dx.doi.org/10.3847/1538-4357/aab9ad}{\JournalTitle{\apj}, 858,
  48}

\bibitem[{{Nagar} {et~al.}(2002){Nagar}, {Oliva}, {Marconi}, \&
  {Maiolino}}]{Nagar2002}
{Nagar}, N.~M., {Oliva}, E., {Marconi}, A., \& {Maiolino}, R. 2002,
  \href{http://dx.doi.org/10.1051/0004-6361:20021039}{\JournalTitle{\aap}, 391,
  L21}

\bibitem[{{Nenkova} {et~al.}(2002){Nenkova}, {Ivezi{\'c}}, \&
  {Elitzur}}]{Nenkova2002}
{Nenkova}, M., {Ivezi{\'c}}, {\v Z}., \& {Elitzur}, M. 2002,
  \href{http://dx.doi.org/10.1086/340857}{\JournalTitle{\apjl}, 570, L9}

\bibitem[{{Osterbrock} \& {Ferland}(2006)}]{Osterbrock2006}
{Osterbrock}, D.~E., \& {Ferland}, G.~J. 2006, {Astrophysics of gaseous nebulae
  and active galactic nuclei}

\bibitem[{{Padovani} {et~al.}(2017){Padovani}, {Alexander}, {Assef}, {De
  Marco}, {Giommi}, {Hickox}, {Richards}, {Smol{\v c}i{\'c}}, {Hatziminaoglou},
  {Mainieri}, \& {Salvato}}]{Padovani2017}
{Padovani}, P., {Alexander}, D.~M., {Assef}, R.~J., {et~al.} 2017,
  \href{http://dx.doi.org/10.1007/s00159-017-0102-9}{\JournalTitle{\aapr}, 25,
  2}

\bibitem[{{Peng} {et~al.}(2006){Peng}, {Gu}, {Melnick}, \& {Zhao}}]{Peng2006}
{Peng}, Z., {Gu}, Q., {Melnick}, J., \& {Zhao}, Y. 2006,
  \href{http://dx.doi.org/10.1051/0004-6361:20054664}{\JournalTitle{\aap}, 453,
  863}

\bibitem[{{Phillips} {et~al.}(1983){Phillips}, {Charles}, \&
  {Baldwin}}]{Phillips1983}
{Phillips}, M.~M., {Charles}, P.~A., \& {Baldwin}, J.~A. 1983,
  \href{http://dx.doi.org/10.1086/160797}{\JournalTitle{\apj}, 266, 485}

\bibitem[{{Polletta} {et~al.}(1996){Polletta}, {Bassani}, {Malaguti},
  {Palumbo}, \& {Caroli}}]{Polletta1996}
{Polletta}, M., {Bassani}, L., {Malaguti}, G., {Palumbo}, G.~G.~C., \&
  {Caroli}, E. 1996,
  \href{http://dx.doi.org/10.1086/192342}{\JournalTitle{\apjs}, 106, 399}

\bibitem[{{Prieto} {et~al.}(2010){Prieto}, {Reunanen}, {Tristram}, {Neumayer},
  {Fernandez-Ontiveros}, {Orienti}, \& {Meisenheimer}}]{Prieto2010}
{Prieto}, M.~A., {Reunanen}, J., {Tristram}, K.~R.~W., {et~al.} 2010,
  \href{http://dx.doi.org/10.1111/j.1365-2966.2009.15897.x}{\JournalTitle{\mnras},
  402, 724}

\bibitem[{{Puccetti} {et~al.}(2004){Puccetti}, {Risaliti}, {Fiore}, {Elvis},
  {Nicastro}, {Perola}, \& {Capalbi}}]{Puccetti2004}
{Puccetti}, S., {Risaliti}, G., {Fiore}, F., {et~al.} 2004,
  \href{http://dx.doi.org/10.1016/j.nuclphysbps.2004.04.039}{\JournalTitle{Nuclear
  Physics B Proceedings Supplements}, 132, 225}

\bibitem[{{Ramos Almeida} {et~al.}(2009){Ramos Almeida}, {P{\'e}rez
  Garc{\'{\i}}a}, \& {Acosta-Pulido}}]{RamosAlmeida2009}
{Ramos Almeida}, C., {P{\'e}rez Garc{\'{\i}}a}, A.~M., \& {Acosta-Pulido},
  J.~A. 2009,
  \href{http://dx.doi.org/10.1088/0004-637X/694/2/1379}{\JournalTitle{\apj},
  694, 1379}

\bibitem[{{Ramos Almeida} {et~al.}(2008){Ramos Almeida}, {P{\'e}rez
  Garc{\'{\i}}a}, {Acosta-Pulido}, \&
  {Gonz{\'a}lez-Mart{\'{\i}}n}}]{RamosAlmeida2008}
{Ramos Almeida}, C., {P{\'e}rez Garc{\'{\i}}a}, A.~M., {Acosta-Pulido}, J.~A.,
  \& {Gonz{\'a}lez-Mart{\'{\i}}n}, O. 2008,
  \href{http://dx.doi.org/10.1086/589771}{\JournalTitle{\apjl}, 680, L17}

\bibitem[{{Ramos Almeida} {et~al.}(2011){Ramos Almeida}, {Levenson},
  {Alonso-Herrero}, {Asensio Ramos}, {Rodr{\'{\i}}guez Espinosa}, {P{\'e}rez
  Garc{\'{\i}}a}, {Packham}, {Mason}, {Radomski}, \&
  {D{\'{\i}}az-Santos}}]{RamosAlmeida2011}
{Ramos Almeida}, C., {Levenson}, N.~A., {Alonso-Herrero}, A., {et~al.} 2011,
  \href{http://dx.doi.org/10.1088/0004-637X/731/2/92}{\JournalTitle{\apj}, 731,
  92}

\bibitem[{{Reunanen} {et~al.}(2002){Reunanen}, {Kotilainen}, \&
  {Prieto}}]{Reunanen2002}
{Reunanen}, J., {Kotilainen}, J.~K., \& {Prieto}, M.~A. 2002,
  \href{http://dx.doi.org/10.1046/j.1365-8711.2002.05181.x}{\JournalTitle{\mnras},
  331, 154}

\bibitem[{{Reunanen} {et~al.}(2003){Reunanen}, {Kotilainen}, \&
  {Prieto}}]{Reunanen2003}
---. 2003,
  \href{http://dx.doi.org/10.1046/j.1365-8711.2003.06771.x}{\JournalTitle{\mnras},
  343, 192}

\bibitem[{{Riffel} {et~al.}(2006){Riffel}, {Rodr{\'{\i}}guez-Ardila}, \&
  {Pastoriza}}]{Riffel2006}
{Riffel}, R., {Rodr{\'{\i}}guez-Ardila}, A., \& {Pastoriza}, M.~G. 2006,
  \href{http://dx.doi.org/10.1051/0004-6361:20065291}{\JournalTitle{\aap}, 457,
  61}

\bibitem[{{Risaliti} {et~al.}(2005){Risaliti}, {Elvis}, {Fabbiano}, {Baldi}, \&
  {Zezas}}]{Risaliti2005}
{Risaliti}, G., {Elvis}, M., {Fabbiano}, G., {Baldi}, A., \& {Zezas}, A. 2005,
  \href{http://dx.doi.org/10.1086/430252}{\JournalTitle{\apjl}, 623, L93}

\bibitem[{{Risaliti} {et~al.}(2002){Risaliti}, {Elvis}, \&
  {Nicastro}}]{Risaliti2002}
{Risaliti}, G., {Elvis}, M., \& {Nicastro}, F. 2002,
  \href{http://dx.doi.org/10.1086/324146}{\JournalTitle{\apj}, 571, 234}

\bibitem[{{Rodr{\'{\i}}guez-Ardila} {et~al.}(2011){Rodr{\'{\i}}guez-Ardila},
  {Prieto}, {Portilla}, \& {Tejeiro}}]{Rodriguez-Ardila2011}
{Rodr{\'{\i}}guez-Ardila}, A., {Prieto}, M.~A., {Portilla}, J.~G., \&
  {Tejeiro}, J.~M. 2011,
  \href{http://dx.doi.org/10.1088/0004-637X/743/2/100}{\JournalTitle{\apj},
  743, 100}

\bibitem[{{Ryder} {et~al.}(1996){Ryder}, {Buta}, {Toledo}, {Shukla},
  {Staveley-Smith}, \& {Walsh}}]{Ryder1996}
{Ryder}, S.~D., {Buta}, R.~J., {Toledo}, H., {et~al.} 1996,
  \href{http://dx.doi.org/10.1086/177000}{\JournalTitle{\apj}, 460, 665}

\bibitem[{{Shankar}(2009)}]{Shankar2009}
{Shankar}, F. 2009,
  \href{http://dx.doi.org/10.1016/j.newar.2009.07.006}{\JournalTitle{\nar}, 53,
  57}

\bibitem[{{She} {et~al.}(2017){She}, {Ho}, \& {Feng}}]{She2017}
{She}, R., {Ho}, L.~C., \& {Feng}, H. 2017,
  \href{http://dx.doi.org/10.3847/1538-4357/835/2/223}{\JournalTitle{\apj},
  835, 223}

\bibitem[{{Stalevski} {et~al.}(2012){Stalevski}, {Fritz}, {Baes}, {Nakos}, \&
  {Popovi{\'c}}}]{Stalevski2012}
{Stalevski}, M., {Fritz}, J., {Baes}, M., {Nakos}, T., \& {Popovi{\'c}}, L.~{\v
  C}. 2012,
  \href{http://dx.doi.org/10.1111/j.1365-2966.2011.19775.x}{\JournalTitle{\mnras},
  420, 2756}

\bibitem[{{Tanaka} {et~al.}(1989){Tanaka}, {Hasegawa}, {Hayashi}, {Brand}, \&
  {Gatley}}]{Tanaka89}
{Tanaka}, M., {Hasegawa}, T., {Hayashi}, S.~S., {Brand}, P.~W.~J.~L., \&
  {Gatley}, I. 1989,
  \href{http://dx.doi.org/10.1086/167006}{\JournalTitle{\apj}, 336, 207}

\bibitem[{{Thatte} {et~al.}(1997){Thatte}, {Quirrenbach}, {Genzel}, {Maiolino},
  \& {Tecza}}]{Thatte1997}
{Thatte}, N., {Quirrenbach}, A., {Genzel}, R., {Maiolino}, R., \& {Tecza}, M.
  1997, \href{http://dx.doi.org/10.1086/304848}{\JournalTitle{\apj}, 490, 238}

\bibitem[{{Tremaine} {et~al.}(2002){Tremaine}, {Gebhardt}, {Bender}, {Bower},
  {Dressler}, {Faber}, {Filippenko}, {Green}, {Grillmair}, {Ho}, {Kormendy},
  {Lauer}, {Magorrian}, {Pinkney}, \& {Richstone}}]{Tremaine2002}
{Tremaine}, S., {Gebhardt}, K., {Bender}, R., {et~al.} 2002,
  \href{http://dx.doi.org/10.1086/341002}{\JournalTitle{\apj}, 574, 740}

\bibitem[{{Xu} {et~al.}(1999){Xu}, {Livio}, \& {Baum}}]{Xu1999}
{Xu}, C., {Livio}, M., \& {Baum}, S. 1999,
  \href{http://dx.doi.org/10.1086/301007}{\JournalTitle{\aj}, 118, 1169}

\end{thebibliography}

\end{document}